\documentclass[epjST]{svjour}

\usepackage{graphicx}
\usepackage{amsmath,bbm,slashed}
\usepackage[latin1]{inputenc}

\newcommand{\calL}{{\cal L}}
\newcommand{\calM}{{\cal M}}
\newcommand{\calO}{{\cal O}}
\newcommand{\SU}{\text{SU}}
\newcommand{\Tr}{\text{Tr}}
\newcommand{\gA}{\texttt{g}_A}
\newcommand{\MS}{\widetilde{\text{MS}}}
\newcommand{\logM}{\ln\frac{M}{\mu}}
\newcommand{\logMsq}{\ln^2\frac{M}{\mu}}

\begin{document}

\title{Chiral effective field theories of the strong interactions}

\author{Matthias R.~Schindler\inst{1}\fnmsep\thanks{\email{schindler@sc.edu}}  \and Stefan Scherer\inst{2}\fnmsep\thanks{\email{scherer@kph.uni-mainz.de}}}

\institute{Department of Physics and Astronomy, University of South Carolina, Columbia, SC 29208, USA \and Institut f\"ur Kernphysik, Johannes Gutenberg-Universit\"at, D-55099 Mainz, Germany}

\abstract{Effective field theories of the strong interactions based on the approximate chiral symmetry of QCD provide a model-independent approach to low-energy hadron physics. We give a brief introduction to mesonic and baryonic chiral perturbation theory and discuss a number of applications. We also consider the effective field theory including vector and axial-vector mesons.}

\maketitle

\section{Introduction}

Effective field theories (EFTs) have become a widely-used tool in a variety of applications such as condensed-matter, nuclear, and particle physics (see, e.g., Refs.~\cite{Polchinski:1992ed,Georgi:1994qn,Kaplan:1995uv,Manohar:1996cq,Pich:1998xt,Burgess:1998ku,Ecker:2005ny,Epelbaum:2005pn,Kaplan:2005es,Donoghue:2009mn,Schindler:2011LNP}). The unifying feature of EFTs is that they are systematic approximations to underlying theories, valid in certain energy domains. In particular, an EFT is applicable for energies that are small compared to an intrinsic energy scale of the underlying theory. While this limits the scope of applicability, it can result in substantial simplifications of calculations or can make calculations possible if it is not known how to apply the underlying theory. Since the focus of the EFT is a specific energy region, it can utilize those degrees of freedom that are relevant to the considered domain, which might in general be different from the degrees of freedom of the underlying theory. An example discussed below is chiral perturbation theory, in which hadronic degrees of freedom such as pions and nucleons are used instead of the more fundamental quarks and gluons. The starting point for the construction of an EFT is a theorem by Weinberg \cite{Weinberg:1978kz}:
\begin{quote}
\ldots if one writes down the most general possible Lagrangian, including {\it all} terms consistent with assumed symmetry principles, and then calculates matrix elements with this Lagrangian to any given order in perturbation theory, the result will simply be the most general possible S-matrix consistent with analyticity, perturbative unitarity, cluster decomposition and the assumed symmetry principles.
\end{quote}
The most general Lagrangian contains an infinite number of terms and therefore does not lend itself to efficient calculations without further simplifications. These can be achieved by application of a so-called ``power counting,'' a scheme that organizes (renormalized) Feynman diagrams according to the relative size of their contributions to physical observables. This ordering in general corresponds to an expansion in the ratio of two scales, $q/\Lambda$. The light scale $q$ is given by the typical momenta for the processes under consideration and the masses of the effective degrees of freedom, while $\Lambda$ corresponds to a heavy scale intrinsic to the underlying theory, such as the mass of a particle not explicitly considered. The power counting can then be used to specify which finite number of terms in the most general Lagrangian are required for a given accuracy of the calculation. Each term in the Lagrangian contains a so-called low-energy constant (LEC). These couplings contain all the physics not explicitly accounted for in the EFT, e.g.,~the contributions from those degrees of freedom that do no appear in the EFT. The LECs cannot be predicted from the EFT, but either have to be calculated from the underlying theory (if feasible) or, more commonly, have to be extracted from experiment. Once determined, the LECs can be used to predict other physical observables.

\section{Chiral perturbation theory}

One of the best-known examples of an EFT is chiral perturbation theory (ChPT), the effective field theory of the strong interactions for energies far below $1\,\text{GeV}$ \cite{Gasser:1983yg,Gasser:1984gg}. Besides the discrete symmetries $C$, $P$, and $T$, quantum chromodynamics (QCD), the theory underlying the strong interactions, exhibits an additional symmetry if the light quark masses are set to zero.  In that case, the QCD Lagrangian is invariant under $\SU(N)_L\times \SU(N)_R$ transformations, where $N$ is the number of massless quarks and the subscript $L$ ($R$) refers to left-handed (right-handed) quark fields.\footnote{There is an additional $\text{U(1)}_V$ invariance related to baryon number conservation, while invariance under $\text{U(1)}_A$ is broken due to an anomaly.} The masses of the up, down, and, to a lesser extent, strange quarks are small compared to the typical hadronic mass scale of roughly $1\,\text{GeV}$. Therefore, the QCD Lagrangian in the chiral limit, i.e., with vanishing light quark masses, might serve as a reasonable starting point to the description of strong-interaction phenomena. Neglecting heavy quarks, the QCD Lagrangian in the chiral limit is given by
\begin{equation}
\calL^0_\text{QCD} = \sum_{l=u,d,s} \left( \bar{q}_{R,l} i \slashed{D}q_{R,l} + \bar{q}_{L,l} i \slashed{D}q_{L,l}\right)-\frac{1}{4}\mathcal{G}_{a\mu\nu}\mathcal{G}_{a}^{\mu\nu},
\end{equation}
where (omitting color and flavor indices) the handed quark fields are given by
\begin{equation}
q_R = \frac{1}{2} \left( \mathbbm{1}+\gamma_5 \right)q,\quad q_L = \frac{1}{2} \left( \mathbbm{1}-\gamma_5 \right)q .
\end{equation}

While no mathematical proof currently exists, there are strong indications that the chiral symmetry $\SU(3)_L\times\SU(3)_R$ of the Lagrangian is spontaneously broken to an $\SU(3)_V$ symmetry of the ground state. According to the Goldstone theorem, the spontaneous breaking of a continuous global symmetry leads to the existence of massless Goldstone bosons. In the case of spontaneous chiral symmetry breaking one expects an octet of massless pseudoscalar Goldstone bosons, which are identified with the $\pi$, $K$, and $\eta$ mesons and the observed finite meson masses are attributed to the nonzero physical quark masses. The existence of this octet of mesons with masses much smaller than other hadron masses is one of the indications that the chiral symmetry of QCD is spontaneously broken; others are the nonvanishing scalar singlet quark condensate and the absence of parity doubling for hadron multiplets.

In addition to spontaneous chiral symmetry breaking, the finite light quark masses result in an \textit{explicit} breaking of the symmetry. Collecting the light quark masses in the matrix $\calM=\text{diag}(m_u,m_d,m_s)$, the quark mass term in the QCD Lagrangian can be written as
\begin{equation}
\calL_\calM = -\bar{q}\calM q = -\left( \bar{q}_R \calM q_L + \bar{q}_L \calM q_R \right).
\end{equation}
The quark mass matrix mixes left- and right-handed quark fields and therefore breaks chiral symmetry explicitly. 

It is convenient to consider the quark mass term as generated by the coupling of the scalar quark density to an external c-number field. Extending this concept to the pseudoscalar quark density, vector and axial-vector currents, one arrives at the QCD Lagrangian in the presence of external fields \cite{Gasser:1983yg,Gasser:1984gg},
\begin{equation}
\label{eq:QCDExternal}
\calL_\text{QCD} = \calL^0_\text{QCD} + \calL_\text{ext}= \calL^0_\text{QCD} + \bar{q}\gamma^\mu \left( v_\mu + \frac{1}{3}v^{(s)}_\mu+\gamma_5 a_\mu\right)q-\bar{q}\left( s-i\gamma_5 p \right) q,
\end{equation}
where the external fields are color-neutral, Hermitian $3\times 3$ matrices. The quark mass matrix is included in $s=\calM+\ldots$, and the standard QCD Lagrangian is obtained by setting all other contributions to the external fields to zero. One can now define a generating functional
\begin{equation}
\exp(i Z[v,a,s,p]) = \langle 0\vert T\exp\left[i \int d^4x \calL_\text{ext}(x) \right] \vert0\rangle,
\end{equation}
from which QCD Green functions can be obtained by functional derivatives with respect to the external fields. In addition, the chiral Ward identities which encode the relations among various Green functions are automatically fulfilled if the generating functional is invariant under local transformations of the external fields. For Gasser and Leutwyler, this approach served as the starting point for the construction of chiral perturbation theory \cite{Gasser:1983yg,Gasser:1984gg}. 

In order to write down the most general Lagrangian describing the interactions of Goldstone bosons with each other and with external fields, we define the $\SU(N)$ matrix
\begin{equation}
\label{eq:U}
U(x)=\exp\left( i\frac{\phi(x)}{F_0} \right),
\end{equation}
where the Hermitian, traceless matrix $\phi(x)$ contains the Goldstone boson fields. For the $\SU(2)$ case $\phi$ is given by
\begin{equation}
\phi =  \sum_{i=1}^{3}\tau_i \phi_i  \equiv \begin{pmatrix} \pi^0 & \sqrt{2}\pi^+ \\ \sqrt{2}\pi^- & -\pi^0  \end{pmatrix} ,
\end{equation}
while for $N=3$
\begin{equation}
\phi =  \sum_{a=1}^{8}\lambda_a \phi_a \equiv 
\begin{pmatrix}
\pi^0+\frac{1}{\sqrt{3}}\eta & \sqrt{2}\pi^+ & \sqrt{2}K^+ \\ 
\sqrt{2}\pi^- & -\pi^0 +\frac{1}{\sqrt{3}}\eta  & \sqrt{2}K^0 \\
\sqrt{2}K^-   & \sqrt{2}\bar{K}^0 & -\frac{2}{\sqrt{3}}\eta
\end{pmatrix}.
\end{equation}
Under $\SU(N)_L\times\SU(N)_R$ the matrix $U(x)$ transforms as
\begin{equation}
U(x) \mapsto R U(x) L^\dagger,
\end{equation}
with $R,L \in \SU(N)$. In combination with the transformation properties under the discrete symmetries $C,P$, and $T$ one can now construct the most general Lagrangian invariant under chiral transformations. The term with the minimal number of derivatives is given by
\begin{equation}
\calL_\text{eff} = \frac{F_0^2}{4}\Tr \left(\partial_\mu U \partial^\mu U^\dagger\right),
\end{equation}
where $F_0$ is the pion decay constant in the chiral limit (with the physical value $F_\pi=92.42(26)\,\text{MeV}$). The explicit symmetry breaking due to finite quark masses can be incorporated via a series of terms containing increasing powers of $\calM$, with the leading-order contribution given by
\begin{equation}
\calL_\text{s.b.}=\frac{F_0^2 B_0}{2}\Tr\left(\calM U^\dagger + U \calM^\dagger \right).
\end{equation}
The low-energy constant $B_0$ can be related to the chiral quark condensate.\footnote{We note that it is customary to use $F_0, B_0$ for the $\SU(3)$ case and $F,B$ for $\SU(2)$. }
Generalizing to local $\SU(N)_L\times\SU(N)_R$ transformations and coupling to the external fields of Eq.~(\ref{eq:QCDExternal}) leads to the introduction of a covariant derivative
\begin{equation}
D_\mu U = \partial_\mu U - ir_\mu U +i U l_\mu,
\end{equation}
where $r_\mu= v_\mu+ a_\mu$ and $l_\mu = v_\mu- a_\mu$. The most general Lagrangian invariant under local chiral transformations with two derivatives and one power of the quark masses is given by
\begin{equation}
\label{eq:LagLO}
\calL_2 = \frac{F_0^2}{4}\Tr \left[D_\mu U (D^\mu U)^\dagger\right]+\frac{F_0^2}{4}\Tr\left(\chi U^\dagger + U \chi^\dagger \right),
\end{equation}
with $\chi=2B_0(s+ip)$.

Equation~(\ref{eq:LagLO}) contains two of the infinite number of terms in the most general Lagrangian. These terms can be ordered according to a power counting due to Weinberg \cite{Weinberg:1978kz}. The Lagrangian is organized as a series of terms with an increasing number of derivatives and powers of the quark masses,
\begin{equation}
\label{eq:LagEff}
\calL_\text{eff} = \calL_2+\calL_4+\calL_6+\ldots.
\end{equation}
This translates into a dual expansion of physical quantities in powers of momenta and quark masses. The contribution of Feynman diagrams, generated by the interactions of Eq.~(\ref{eq:LagEff}), can be characterized by their chiral order $D$, which is assigned according to
\begin{equation}
\label{eq:PowerCount}
D=2+2N_L +\sum_{n=1}^{\infty} N_{2n} (2n-2),
\end{equation}
with $N_L$ the number of independent loops and $N_{2n}$ the number of vertices originating from $\calL_{2n}$. Equation~(\ref{eq:PowerCount}) establishes a relation between the chiral and the loop expansions. One-loop diagrams start to contribute at $D=4$, while two-loop diagrams can enter at $D=6$. In the following, calculations to $\calO(q^n)$ include diagrams up to and including chiral order $D=n$.

The Lagrangian $\calL_4$ is given in Refs.~\cite{Gasser:1983yg,Gasser:1984gg}. In the $\SU(3)$ case it contains 10 independent terms (not counting contact terms without Goldstone boson fields), while for $\SU(2)$ it consists of 7 terms.  The sixth-order Lagrangian ${\cal L}_6$ was constructed in Refs.~\cite{Fearing:1994ga,Bijnens:1999sh,Ebertshauser:2001nj,Bijnens:2001bb} and contains 90 + 23 terms in the $\SU(3)$ case. While the number of corresponding LECs that have to be determined from experiment might sound large, it is important to keep in mind that only a small subset of these contributes to any given physical observable.

Chiral perturbation theory has been applied to a large number of pseudoscalar meson properties and reactions. In the following we briefly discuss two examples, the masses and electromagnetic polarizabilities. For detailed reviews we refer the reader to Refs.~\cite{Meissner:1993ah,Leutwyler:1994fi,Ecker:1994gg,Pich:1995bw,Scherer:2002tk,Bijnens:2006zp}.

\subsection{Masses of the pseudoscalar mesons}

\begin{figure}[t]
\begin{center}
\resizebox{0.5\textwidth}{!}{%
\includegraphics{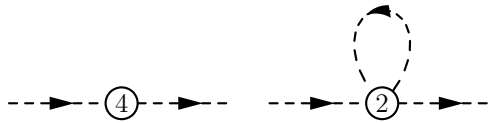}
}
\end{center}
\caption{Self-energy diagrams at ${\cal O}(q^4)$.
   Vertices derived from ${\cal L}_{2n}$ are denoted by $2n$ in the
interaction blobs.}\label{3:3:3:selfenergy}
\end{figure}

   The quark-mass expansion of the Goldstone boson masses at ${\cal O}(q^4)$ is one of the
simplest applications of ChPT beyond tree level.
   For simplicity, we restrict ourselves to the isospin-symmetric limit, $m_u=m_d=\hat m$.
   At ${\cal O}(q^2)$, the terms of second order in the fields in ${\cal L}_{\rm s.b.}$,
\begin{equation}
\label{3:2:2:lmzo}
{\cal L}_{\rm s.b}=-\frac{B_0}{2}\mbox{Tr}(\phi^2{\cal M}) +\ldots,
\end{equation}
   generate the following expressions for the masses of the Goldstone bosons,
\begin{equation}
\label{mgb2}
M^2_{\pi,2}=2 B_0 \hat m,\quad
M^2_{K,2}=B_0(\hat m+m_s),\quad
M^2_{\eta,2}=\frac{2}{3} B_0\left(\hat m+2m_s\right).
\end{equation}
   The subscript 2 refers to chiral order 2.
   Without additional input regarding the numerical value of $B_0$,
Eqs.\ (\ref{mgb2}) do not allow for an extraction of
the  absolute values of the quark masses $\hat m$ and $m_s$, because rescaling
$B_0\to \lambda B_0$ in combination with $m_q\to m_q/\lambda$ leaves the
relations invariant.
   For the {\it ratio} of the quark masses one obtains, using the empirical values
$M_\pi=135$ MeV, $M_K=496$ MeV,
and $M_\eta=548$ MeV,
\begin{align}
\frac{M^2_{K,2}}{M^2_{\pi,2}}=\frac{\hat m+m_s}{2\hat m}&\Rightarrow\frac{m_s}{\hat m}=25.9,&
\frac{M^2_{\eta.2}}{M^2_{\pi,2}}=\frac{2m_s+\hat m}{3\hat m}&\Rightarrow\frac{m_s}{\hat m}=24.3.
\end{align}
    The masses beyond tree level are given by the solutions to the equations
\begin{equation}
\label{mdef}
M^2_\phi-M^2_{\phi,2}-\Sigma_\phi(M^2_\phi)=0,\quad \phi=\pi,K,\eta,
\end{equation}
   corresponding to the poles of the full propagators.
   The proper self-energy insertions, $-i\Sigma_\phi(p^2)$, consist of
one-particle-irreducible diagrams only, i.e., diagrams which do not fall apart into
two separate pieces when cutting an arbitrary internal line.
   The accuracy of the determination of $M^2_\phi$ depends on the accuracy of
the calculation of $\Sigma_\phi$.
   At chiral order $D=4$, the self energies originate from the diagrams
shown in Fig.\ \ref{3:3:3:selfenergy} and are of the form
\begin{equation}
\label{sigmaphi}
\Sigma_{\phi,4}(p^2)=A_\phi+B_\phi p^2,
\end{equation}
   where the constants $A_\phi$ and $B_\phi$ receive a tree-level
contribution from ${\cal L}_4$ and a one-loop contribution with a
vertex from ${\cal L}_2$.
   The masses at ${\cal O}(q^4)$ are determined by solving Eq.~(\ref{mdef})
with the self energies of Eq.\ (\ref{sigmaphi}),
\begin{displaymath}
M^2_{\phi,4}=M_{\phi,2}^2(1+B_\phi)+A_\phi,
\end{displaymath}
   because $A_\phi={\cal O}(q^4)$ and $\{B_\phi, M_{\phi,2}^2\}={\cal O}(q^2)$.
   The results for the masses of the Goldstone bosons at ${\cal O}(q^4)$
read \cite{Gasser:1984gg}
\begin{eqnarray}
\label{3:5:mpi24}
M^2_{\pi,4}&=&M^2_{\pi,2}\Bigg\{1+\frac{M^2_{\pi,2}}{32\pi^2F^2_0}
\ln\left(\frac{M^2_{\pi,2}}{\mu^2}\right)-\frac{M^2_{\eta,2}}{96\pi^2F^2_0}
\ln\left(\frac{M^2_{\eta,2}}{\mu^2}\right)\nonumber\\
&&+\frac{16}{F^2_0}\left[(2\hat m+m_s)B_0(2L^r_6-L^r_4)
+\hat mB_0(2L^r_8-L^r_5)\right]\Bigg\},\\
\label{3:5:mk24}
M^2_{K,4}&=&M^2_{K,2}\Bigg\{1+\frac{M^2_{\eta,2}}{48\pi^2F^2_0}
\ln\left(\frac{M^2_{\eta,2}}{\mu^2}\right)\nonumber\\
&&+\frac{16}{F^2_0}\left[(2\hat m+m_s)B_0(2L^r_6-L^r_4)
+\frac{1}{2}(\hat m+m_s)B_0(2L^r_8-L^r_5)\right]\Bigg\},\nonumber\\
&&\\
\label{3:5:meta24}
M^2_{\eta,4}&=&M^2_{\eta,2}\left[1+\frac{M^2_{K,2}}{16\pi^2F^2_0}
\ln\left(\frac{M^2_{K,2}}{\mu^2}\right)
-\frac{M^2_{\eta,2}}{24\pi^2F^2_0}\ln\left(\frac{M^2_{\eta,2}}{\mu^2}\right)
\right.\nonumber\\
&&\left.+\frac{16}{F^2_0}(2\hat m+m_s)B_0(2L^r_6-L^r_4)
+8\frac{M^2_{\eta,2}}{F^2_0}(2L^r_8-L^r_5)\right]\nonumber\\
&&+M^2_{\pi,2}\left[\frac{M^2_{\eta,2}}{96\pi^2F^2_0}
\ln\left(\frac{M^2_{\eta,2}}{\mu^2}\right)
-\frac{M^2_{\pi,2}}{32\pi^2F^2_0}
\ln\left(\frac{M^2_{\pi,2}}{\mu^2}\right)
+\frac{M^2_{K,2}}{48\pi^2F^2_0}
\ln\left(\frac{M^2_{K,2}}{\mu^2}\right)\right]\nonumber\\
&&+\frac{128}{9}\frac{B^2_0(\hat m-m_s)^2}{F^2_0}
(3L^r_7+L^r_8).
\end{eqnarray}
The above results for the masses exhibit a few characteristic properties of ChPT.
   The loop diagrams of Fig.~\ref{3:3:3:selfenergy} generate ultraviolet
divergences at ${\cal O}(q^4)$ which are absorbed by an appropriate renormalization of
the LECs $L_i$ of the most general Lagrangian ${\cal L}_4$.
   Therefore, the expressions for the masses are finite.
   By construction, the dependence of the renormalized coefficients $L_i^r$ on the
renormalization scale $\mu$ is such that it cancels the
scale dependence of the logarithms in Eqs.\ (\ref{3:5:mpi24}) -
(\ref{3:5:meta24}).
   Thus, physical observables do not depend on the scale $\mu$.
   At ${\cal O}(q^4)$, the masses of the Goldstone bosons
vanish if the quark masses are sent to zero.
   This is, of course, what we expected from QCD in the chiral limit
but it is reassuring to verify that the self interaction in ${\cal L}_2$
(in the absence of quark masses) does not generate
Goldstone-boson masses at higher order.
The expressions of Eqs.~(\ref{3:5:mpi24}) - (\ref{3:5:meta24}) contain two types of contributions at $\calO(q^4)$. Terms that are analytic in the quark masses, i.e., proportional to $m_q^2$, are multiplied by the renormalized LECs $L_i^r$ of $\calL_4$. However, the terms of the type $m^2_q \ln(m_q)$---so-called  chiral logarithms---are nonanalytic in the quark masses and do not involve new parameters.
   Such a behavior is an illustration of the mechanism found by Li and
Pagels \cite{Li:1971vr}, who noticed that a perturbation theory
around a symmetry which is realized in the Nambu-Goldstone mode
results in both analytic as well as nonanalytic expressions in
the perturbation.

\subsection{Electromagnetic polarizabilities of the pion} \label{sec:2}

In the framework of classical electrodynamics, the electric and magnetic polarizabilities $\alpha$ and $\beta$ describe the response of a system to static, uniform, external electric and magnetic fields in terms of induced electric and magnetic dipole moments. In principle, empirical information on the pion polarizabilities can be obtained from the differential cross section of low-energy Compton scattering on a charged pion, $\gamma(\omega,\vec q)+\pi^+(E_\pi,\vec p)\to\gamma(\omega',\vec q')+\pi^+(E_\pi',\vec p')$, 
\begin{align*}
\frac{d\sigma}{d\Omega_\text{lab}}&= \left(\frac{\omega'}{\omega}\right)^2 \frac{e^2}{4\pi M_\pi}\left\{\frac{e^2}{4\pi M_\pi}\frac{1+z^2}{2}\right.\\
&\quad\left.-\frac{\omega\omega'}{2} \left[(\alpha+\beta)_{\pi^+}(1+z)^2 +(\alpha-\beta)_{\pi^+}(1-z)^2\right]\right\} +\ldots,
\end{align*}
where $e$ is the elementary charge, $M_\pi$ the pion mass, $z=\hat{q}\cdot\hat{q}\,'$, and $\omega'/\omega=[1+\omega(1-z)/M_\pi]$. The forward and backward differential cross sections are sensitive to $(\alpha+\beta)_{\pi^+}$ and $(\alpha-\beta)_{\pi^+}$, respectively.

The predictions for the charged pion polarizabilities at ${\cal O}(q^4)$ \cite{Bijnens:1987dc},
\begin{displaymath}
\alpha_{\pi^+}=-\beta_{\pi^+}=2 \frac{e^2}{4\pi} \frac{1}{(4\pi F_\pi)^2 M_\pi}\frac{\bar l_6-\bar l_5}{6} =(2.64\pm 0.09) \times 10^{-4}\, \mbox{fm}^3,
\end{displaymath}
correspond to an old current-algebra low-energy theorem \cite{Terentev:1972ix} which relates Compton scattering on a charged pion to radiative charged-pion beta decay, $\pi^+\to e^+\nu_e\gamma$, by means of a chiral Ward identity.
At ${\cal O}(q^4)$, the linear combination ${\bar{l}}_\Delta\equiv\bar l_6-\bar l_5$ of scale-independent LECs \cite{Gasser:1983yg} is related to the ratio $\gamma=F_A/F_V$ of the pion axial-vector form factor $F_A$ and the vector form factor $F_V$ of radiative pion beta decay~\cite{Gasser:1983yg}, $\gamma={\bar{l}}_\Delta/6$. Once this ratio is known, chiral symmetry makes an {\it absolute} prediction for the polarizabilities. This situation is similar to the $s$-wave $\pi\pi$-scattering lengths at lowest order, ${\cal O}(q^2)$, which are predicted once $F_\pi$ is known. Using the most recent determination $\gamma=0.443\pm 0.015$ by the PIBETA Collaboration~\cite{Frlez:2003pe} (assuming $F_V=0.0259$ obtained from the conserved vector current hypothesis) results in the ${\cal O}(q^4)$ prediction $\alpha_{\pi^+}=(2.64\pm 0.09)\times 10^{-4}\, \mbox{fm}^3$, where the error estimate is only due to the error of $\gamma$ and does not include effects from higher orders in the quark-mass expansion.

Corrections to the leading-order result have been calculated at ${\cal O}(q^6)$ and turn out to be rather small~\cite{Burgi:1996qi,Gasser:2006qa}. Using updated values for the LECs, the predictions of Ref.~\cite{Gasser:2006qa} are
\begin{eqnarray}
(\alpha + \beta)_{\pi^+} &=& 0.16 \times 10^{-4}\, \mbox{fm}^3,\label{eq:1.1}\\
(\alpha - \beta)_{\pi^+} &=& (5.7 \pm 1.0)\times 10^{-4}\, \mbox{fm}^3.\label{eq:1.2}
\end{eqnarray} 
The degeneracy $\alpha_{\pi^+}=-\beta_{\pi^+}$ is lifted at the two-loop level. The corresponding corrections to the ${\cal O}(q^4)$ result indicate a similar rate of convergence as for the $\pi\pi$-scattering lengths \cite{Gasser:1983yg,Bijnens:1995yn}. The error for $(\alpha + \beta)_{\pi^+}$ is of the order $0.1\times 10^{-4}\, \mbox{fm}^3$, mostly from the dependence on the scale at which the ${\cal O}(q^6)$ LECs are estimated by resonance saturation.

As there is no stable pion target, empirical information on the pion polarizabilities is not easily obtained. For that purpose, one has to consider reactions which contain the Compton scattering amplitude as a building block, such as, e.g., the Primakoff effect in high-energy pion-nucleus bremsstrahlung, $\pi^-Z\to \pi^-Z \gamma$ \cite{Antipov:1982kz}, radiative pion photoproduction on the nucleon, $\gamma p\to \gamma \pi^+n$ \cite{Aibergenov:1986gi,Ahrens:2004mg}, and pion pair production in $e^+e^-$ scattering, $e^+e^-\to e^+e^-\pi^+\pi^-$ \cite{Berger:1984xb,Courau:1986gn,Ajaltoni,Boyer:1990vu}. The results of the older experiments are summarized in Table \ref{tab:pionpolexp}. Recently, also the COMPASS Collaboration at CERN has investigated the Primakoff reaction, and the data analysis is underway \cite{Guskov:2008zz}.

\begin{table}
\caption{Experimental data on the charged pion polarizability $\alpha_{\pi^+}$}
\label{tab:pionpolexp}
\begin{center}
\begin{tabular}{llc}
\hline\noalign{\smallskip} Reaction&Experiment&$\alpha_{\pi^+}$ [$10^{-4}$ $\mbox{fm}^3$]\\
\noalign{\smallskip}\hline\noalign{\smallskip} $\pi^{-}Z\rightarrow
\pi^{-}Z\gamma$ & Serpukhov \cite{Antipov:1982kz}& $6.8\pm 1.4\pm 1.2$\\
$\gamma p\to \gamma\pi^+ n$& Lebedev Phys.~Inst.~\cite{Aibergenov:1986gi}&$20\pm 12$\\
$\gamma \gamma \rightarrow \pi^+\pi^-$ & PLUTO \cite{Berger:1984xb}& $19.1\pm 4.8\pm 5.7$\\
& DM 1 \cite{Courau:1986gn}& $17.2\pm 4.6$      \\
& DM 2 \cite{Ajaltoni}& $26.3\pm 7.4$      \\
& MARK II \cite{Boyer:1990vu}& $2.2\pm 1.6$     \\
\noalign{\smallskip}\hline
\end{tabular}
\end{center}
\end{table}

\begin{figure}[t]
\begin{center}
\resizebox{0.3\textwidth}{!}{%
\includegraphics{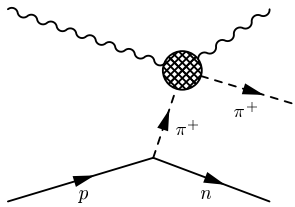}
}
\end{center}
\caption{The reaction $\gamma p\to\gamma\pi^+n$ contains Compton scattering on a pion as a sub diagram in the $t$ channel, where $t=(p_n-p_p)^2$.}
\label{fig:tchannel}
\end{figure}

\begin{figure}[t]
\begin{minipage}[t]{0.45\textwidth}
\begin{center}
\resizebox{1\textwidth}{!}{%
\includegraphics{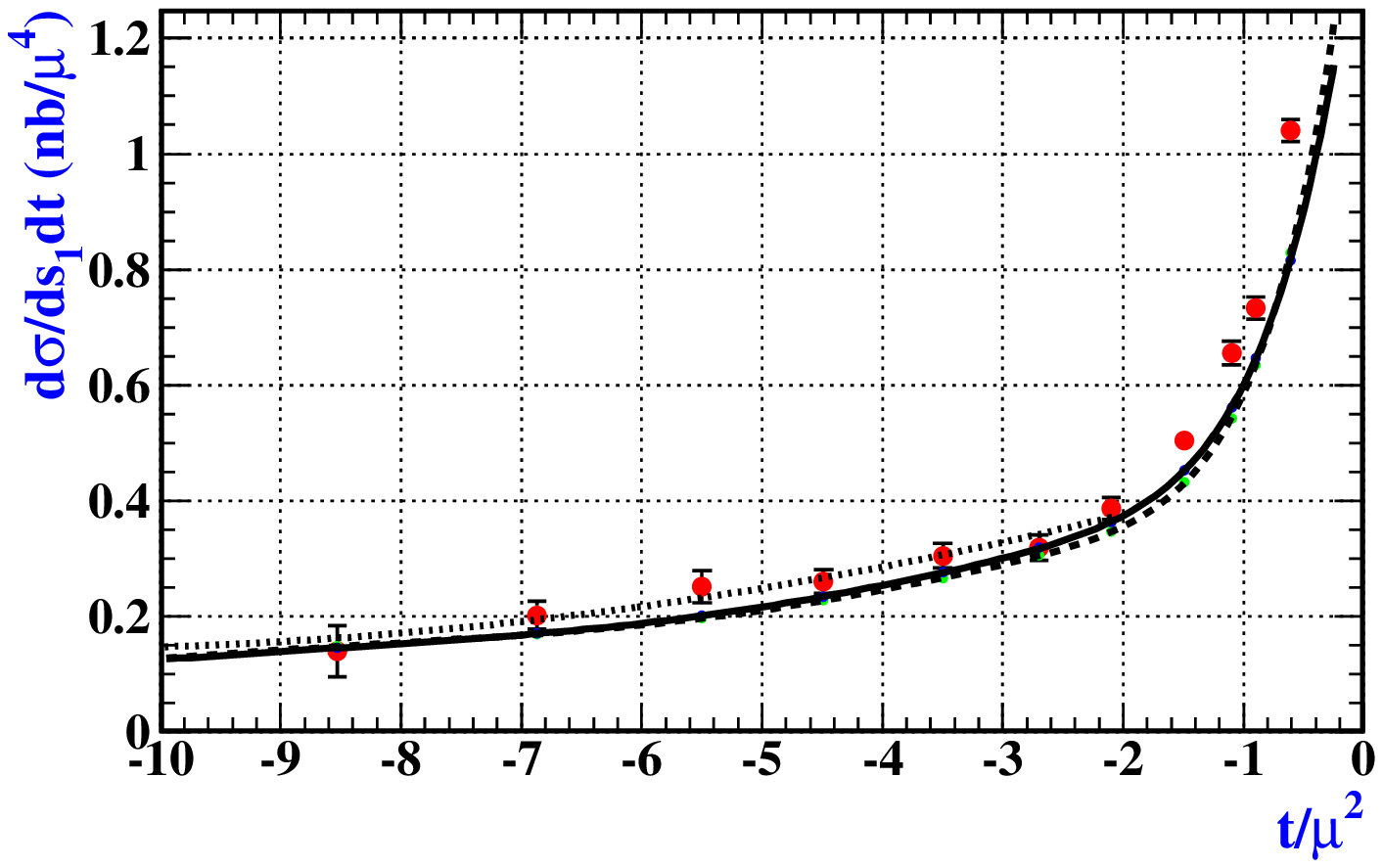}
}
\end{center}
\caption{Differential cross section averaged over 537 MeV $< E_\gamma <$ 817 MeV and 1.5 $M_\pi^2<s_1<5 M_\pi^2$. Solid line: model 1; dashed line: model 2; dotted line: fit to experimental data.} \label{fig:cros1c}
\end{minipage}
\hspace{0.1\textwidth}
\begin{minipage}[t]{0.45\textwidth}
\begin{center}
\resizebox{\textwidth}{!}{%
\includegraphics{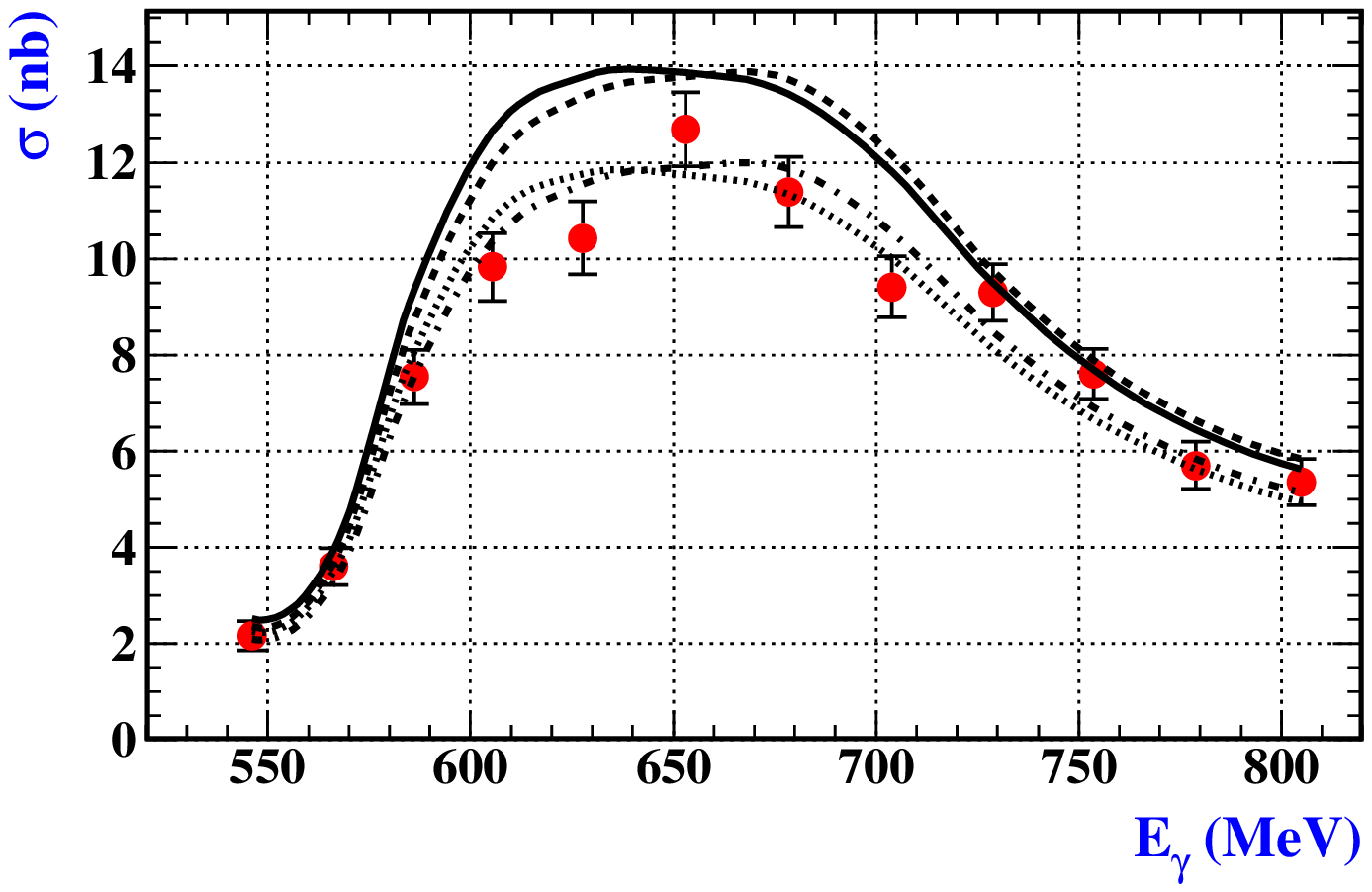}
}
\end{center}
\caption{Cross section of $\gamma p\to\gamma\pi^+n$ integrated over $s_1$ and $t$ in the region where the contribution of the pion polarizability is biggest and the difference between the predictions of the theoretical models under consideration does not exceed 3 \%. The dashed and dashed-dotted lines are predictions of model 1 and the solid and dotted lines of model 2 for $(\alpha-\beta)_{\pi^+}=0\,\text{fm}^3$ and $(\alpha-\beta)_{\pi^+}=14\times 10^{-4}\,
\mbox{fm}^3$, respectively.} \label{fig:cros2c}
\end{minipage}
\end{figure}

The potential of obtaining information on the pion polarizabilities from radiative pion photoproduction from the proton was extensively studied in Ref.~\cite{Drechsel:1994kh}. In terms of Feynman diagrams, the reaction $\gamma p\to\gamma\pi^+n$ contains real Compton scattering on a charged pion as a pion pole diagram (see Fig.~\ref{fig:tchannel}). In the most recent experiment on $\gamma p\to\gamma\pi^+n$ at the Mainz Microtron MAMI \cite{Ahrens:2004mg}, the cross section was obtained in the kinematic region $537\,\text{MeV} < E_\gamma < 817\,\text{MeV}$, $140^{\circ}\le\theta^{\rm cm}_{\gamma\gamma'}\leq 180^{\circ}$. Instead of performing an extrapolation to the $t$-channel pole of the Chew-Low type \cite{Chew:1958wd,Unkmeir}, the values of the pion polarizabilities were obtained from the data by a fit of the cross section as calculated by different theoretical models. Figure \ref{fig:cros1c} shows the experimental data, averaged over the full photon beam energy interval and over the squared pion-photon center-of-mass energy $s_1$ from $1.5\,M_\pi^2$ to  $5\,M_\pi^2$, as a function of the squared pion momentum transfer $t$ in units of $M_\pi^2$. For such small values of $s_1$, the differential cross section is expected to be insensitive to the pion polarizabilities. Also shown are two model calculations: model 1 (solid curve) is a simple Born approximation using the pseudoscalar pion-nucleon interaction including the anomalous magnetic moments of the nucleon; model 2 (dashed curve) consists of pole terms without the anomalous magnetic moments but including contributions from the resonances $\Delta (1232)$, $P_{11}(1440)$, $D_{13}(1520)$ and $S_{11}(1535)$. The dotted curve is a fit to the experimental data.

The kinematic region where the polarizability contribution is biggest is given by $5M_\pi^2< s_1<15M_\pi^2$ and $-12M_\pi^2<t<-2M_\pi^2$. Figure~\ref{fig:cros2c} shows the cross section as a function of the beam energy integrated over $s_1$ and $t$ in this second region. The dashed (dashed-dotted) and solid (dotted)  lines refer to models 1 and 2, respectively, each with $(\alpha-\beta)_{\pi^+}$ set equal to $0\, \mbox{fm}^3$  ($14\times 10^{-4}\, \mbox{fm}^3$). By comparing the 12 experimental points with the predictions of the models, the corresponding values of $(\alpha-\beta)_{\pi^+}$ for each data point were determined in combination with the corresponding statistical and systematic errors. The result extracted from the combined analysis of the 12 data points reads \cite{Ahrens:2004mg}
\begin{equation} 
\label{alphambeta} (\alpha-\beta)_{\pi^+}=(11.6\pm 1.5_{\rm stat}\pm 3.0_{\rm syst}\pm 0.5_{\rm mod}) \times 10^{-4}\, \mbox{fm}^3
\end{equation}
and has to be compared with the ChPT result of, e.g., Ref.~\cite{Gasser:2006qa},  $(5.7\pm 1.0) \times 10^{-4}\, \mbox{fm}^3$ which deviates by 2 standard deviations from the experimental result.

Clearly, the model-dependent input to the result of Eq.~(\ref{alphambeta}) deserves further study. In particular, the model error was estimated by comparing the analysis with two specific models. In Ref.\ \cite{Kao:2008pf} radiative pion photoproduction was studied in the framework of heavy-baryon chiral perturbation theory (see Sec.~\ref{sec:BChPT}) at the one-loop level. Unfortunately, the kinematical conditions of the MAMI experiment were not explicitly considered. It was argued that the extraction of pion polarizabilities is, in principle, possible and that the main uncertainty in the extraction arises from the effect of two structures of the ${\cal O}(q^3)$ Lagrangian.

In addition to the apparent disagreement with data, there has been a long-standing problem on the theoretical side. The application of dispersion sum rules as performed in \cite{Fil'kov:1998np,Fil'kov:2005ss} yields $(\alpha-\beta)_{\pi^+}=(13.0^{+2.6}_{-1.9}) \times 10^{-4}\, \mbox{fm}^3$ which provides an even more pronounced discrepancy with the predictions of chiral perturbation theory than the MAMI result \cite{Gasser:2006qa}. These dispersion relations are based on specific forms for the absorptive part of the Compton amplitudes. In Ref.\ \cite{Pasquini:2008ep}, the analytic properties of these forms have been examined and the strong enhancement of intermediate-meson contributions was shown to be connected with spurious singularities.

Clearly, the electromagnetic polarizabilities of the charged pion remain one of the challenging topics of hadronic physics in the low-energy domain. Chiral symmetry provides a strong constraint in terms of radiative pion beta decay and mesonic chiral perturbation theory makes a firm prediction beyond the current-algebra result at the two-loop level. Both the experimental determination as well as the theoretical extraction from experiment require further efforts. For a discussion of the so-called generalized pion polarizabilities see Refs.~\cite{Unkmeir:1999md,Fuchs:2000pn,L'vov:2001fz,Unkmeir:2001gw}.

\section{Baryon chiral perturbation theory}\label{sec:BChPT}

The extension of chiral perturbation theory to include the interaction with baryon fields was first discussed in Ref.~\cite{Gasser:1987rb}, considering matrix elements with one-baryon initial and final states. The objective is the description of baryon properties such as masses and form factors as well as, e.g., scattering processes at low energies. The starting point is again the most general Lagrangian, for which the transformation properties of the baryon fields under chiral transformations need to be known. In the following we consider the $\SU(2)\times\SU(2)$ case. For the generalization to  $\SU(3)\times\SU(3)$ see, e.g., Ref.~\cite{Krause:1990xc}. The nucleon fields are collected in
\begin{equation}
\Psi=\begin{pmatrix} p \\ n \end{pmatrix},
\end{equation}
where $p$ and $n$ are the four-component Dirac fields for the proton and neutron, respectively. The nucleons transforms under $\SU(2)\times\SU(2)$ as
\begin{equation}
\Psi \mapsto K(L,R,U) \Psi,
\end{equation}
where
\begin{equation}
K(L,R,U) = \sqrt{RUL^\dagger}^{-1} R \sqrt{U}
\end{equation}
with $L,R \in \SU(2)$ and $U$ the Goldstone boson matrix defined in Eq.~(\ref{eq:U}). Introducing the connection \cite{Ecker:1994gg}
\begin{equation}
\Gamma_\mu = \frac{1}{2}\left[ u^\dagger(\partial_\mu-ir_\mu)u + u(\partial_\mu-il_\mu)u^\dagger \right],
\end{equation}
where $u^2=U$, the covariant derivative of the nucleon field is defined as
\begin{equation}
D_\mu\Psi = \left( \partial_\mu +\Gamma_\mu-iv^{(s)}_\mu \right)\Psi.
\end{equation}
It is also convenient to introduce
\begin{displaymath}
u_\mu =i\left[u^\dagger(\partial_\mu-ir_\mu)u-u(\partial_\mu-il_\mu)u^\dagger\right].
\end{displaymath}
The lowest-order Lagrangian is given by \cite{Gasser:1987rb}
\begin{equation}
\calL^{(1)}_{\pi N} = \bar{\Psi}\left(i\slashed{D} - m + \frac{\gA}{2}\gamma^\mu \gamma_5 u_\mu\right)\Psi, 
\end{equation}
containing two LECs: $m$, the nucleon mass in the chiral limit, and $\gA$, the nucleon axial-vector coupling constant in the chiral limit. We denote the physical values of these parameters as $m_N$ and $g_A$, respectively. Unlike in the purely mesonic sector, the baryonic effective Lagrangian contains contributions at even and odd orders. The Lagrangians up to order $q^4$ are given in Ref.~\cite{Fettes:2000gb}.

The assignment of specific chiral orders to terms in the Lagrangian assumes the existence of a consistent power counting. When the methods of mesonic ChPT were first applied to the one-nucleon sector, however, it was noted that loop diagrams contributed to lower orders than predicted by the power counting \cite{Gasser:1987rb}. As explained in Ref.~\cite{Gasser:1987rb}, the nucleon mass in the chiral limit does not vanish and therefore constitutes an additional scale. It was also noted that the violation of the power counting was due to applying dimensional regularization in combination with the modified minimal subtraction scheme of ChPT ($\MS$) to loop diagrams, and that the ``same phenomenon would occur in the meson sector, if one did not make use of dimensional regularization'' \cite{Gasser:1987rb}. Subsequently, multiple solutions to the apparent power-counting problem were proposed (see, e.g., Refs.~\cite{Jenkins:1990jv,Ellis:1997kc,Becher:1999he,Gegelia:1999gf,Gegelia:1999qt,Fuchs:2003qc}).  

In heavy-baryon ChPT (HBChPT) \cite{Jenkins:1990jv}, an additional expansion of the effective Lagrangian in inverse powers of the nucleon mass in the chiral limit similar to a Foldy-Wouthuysen transformation is performed. The application of dimensional regularization and the $\MS$ scheme to loop diagrams results in a consistent power counting as in the mesonic sector. A large number of physical observables has been calculated in this approach (see, e.g., Ref.~\cite{Bernard:1995dp}).

The expansion in inverse powers of $m$ increases the number of terms in the effective Lagrangian and thus the calculational effort especially when going to higher orders. Furthermore, the expansion in some cases creates problems with analyticity \cite{Becher:1999he} as the poles from nucleon propagators are shifted due to the additional expansion. In Ref.~\cite{Becher:1999he} a different solution to the power counting problem, referred to as infrared regularization, was proposed which keeps the analytic properties of low-energy amplitudes intact.  The main idea is to separate loop integrals into an infrared-singular part that contains the same infrared singularities as the original integral, and an infrared-regular part that is analytic in small parameters for any arbitrary number of space-time dimensions. The infrared-regular part can be absorbed in the LECs of the effective theory, and the remaining contributions from the infrared-singular parts obey the power counting. Infrared regularization has become one of the standard approaches to BChPT calculations (see, e.g., Ref.~\cite{Bernard:2007zu} for an overview).

A different approach to a manifestly Lorentz-invariant formulation of BChPT is given by the extended on-mass-shell (EOMS) scheme \cite{Fuchs:2003qc}. It provides a method to absorb in the LECs of the theory exactly those terms that violate the power counting. While in its original formulation the infrared regularization of Ref.~\cite{Becher:1999he} is applicable to one-loop diagrams containing nucleon and pion propagators, the EOMS scheme can also be applied to multi-loop diagrams \cite{Schindler:2003je} and diagrams containing other degrees of freedom \cite{Fuchs:2003sh}. It was subsequently realized that the infrared regularization can be formulated similarly to the EOMS scheme \cite{Schindler:2003xv}, which extends the applicability to multi-loop and heavy-meson diagrams \cite{Schindler:2003je}. A different extension of infrared regularization is given in Ref.~\cite{Bruns:2004tj}.

As in the mesonic sector, we only discuss a few applications of BChPT in the following. Detailed reviews can be found, e.g., in Refs.~\cite{Bernard:1995dp,Bernard:2007zu,Scherer:2009bt}.

\subsection{Nucleon mass to $\calO(q^6)$}\label{subsec:NucMass}

The nucleon mass $m_N$ provides a good testing ground for applications of BChPT, as it does not depend on any momentum transfers and the chiral expansion therefore corresponds to an expansion in the quark masses. For this reason, the calculation of the nucleon mass has been performed in all renormalization schemes mentioned above \cite{Gasser:1987rb,Bernard:1992qa,McGovern:1998tm,Becher:1999he,Fuchs:2003qc}. With the exception of Ref.~\cite{Gasser:1987rb}, these schemes have in common that they establish the connection between the chiral and the loop expansions, analogous to the mesonic sector. A calculation that only includes one-loop diagrams can be performed up to and including $\calO(q^4)$. The general form of the chiral expansion is given by
\begin{equation}
\label{eq:NucMass4}
m_N = m + k_1 M^2 + k_2 M^3 + k_3 M^4 \logM +k_4 M^4 + \ldots,
\end{equation}
where $M^2=2B\hat m$ is the lowest-order expression for the squared pion mass, the ellipsis stands for higher-order terms, and $\mu$ is a renormalization scale. As an example, in the EOMS scheme the expressions for the $k_i$ are given by \cite{Fuchs:2003qc}
\begin{equation}
\begin{split}
k_1 &= -4c_1,\quad k_2 = -\frac{3\gA^2}{32\pi F^2} , \quad k_3 = \frac{3}{32\pi^2 F^2} \left( 8c_1-c_2-4c_3-\frac{\gA^2}{m} \right),\\
k_4 &= \frac{3\gA^2}{32\pi^2 F^2 m}(1+4c_1 m) + \frac{3}{128\pi^2 F^2}c_2-2(8e_{38}+e_{115}+e_{116}).
\end{split}
\end{equation}
Here, the $c_i$ and $e_j$ are LECs of the second- and fourth-order baryonic Lagrangians, respectively. The expression of Eq.~(\ref{eq:NucMass4}) together with estimates of the various LECs was used in Ref.~\cite{Fuchs:2003kq} to determine the nucleon mass in the chiral limit,
\begin{align}
m &= m_N -\Delta m \notag\\
&= [938.3 - 74.8+15.3+4.7+1.6-2.3]\, \text{MeV}\\
&= 882.8\,\text{MeV},\notag
\end{align}
i.e., $\Delta m = 55.5\,\text{MeV}$.
\begin{figure}[t]
\begin{minipage}{0.45\textwidth}
\begin{center}
\resizebox{1\textwidth}{!}{
\includegraphics{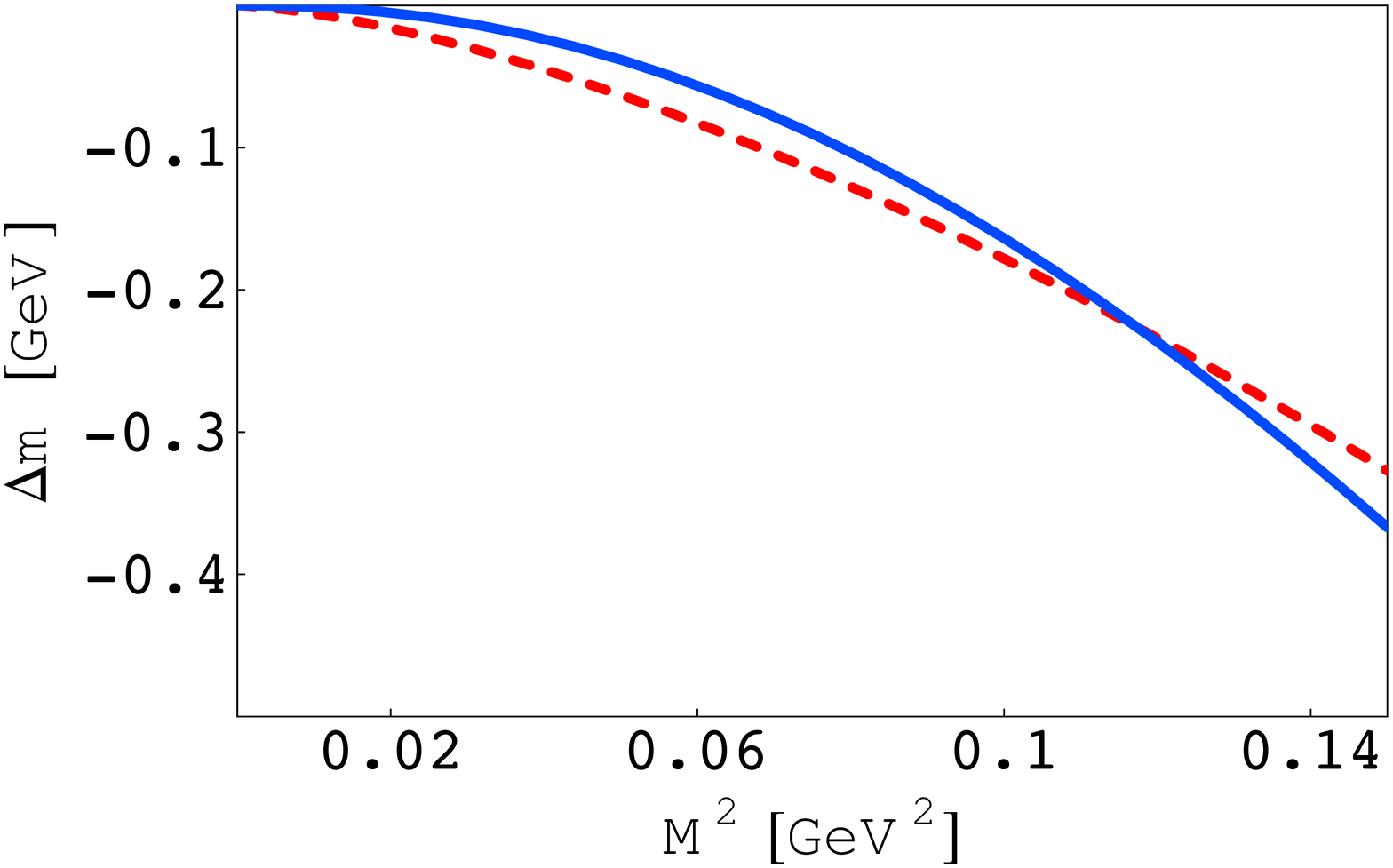}
}
\end{center}
\end{minipage}
\hspace{0.1\textwidth}
\begin{minipage}{0.45\textwidth}
\begin{center}
\resizebox{\textwidth}{!}{
\includegraphics{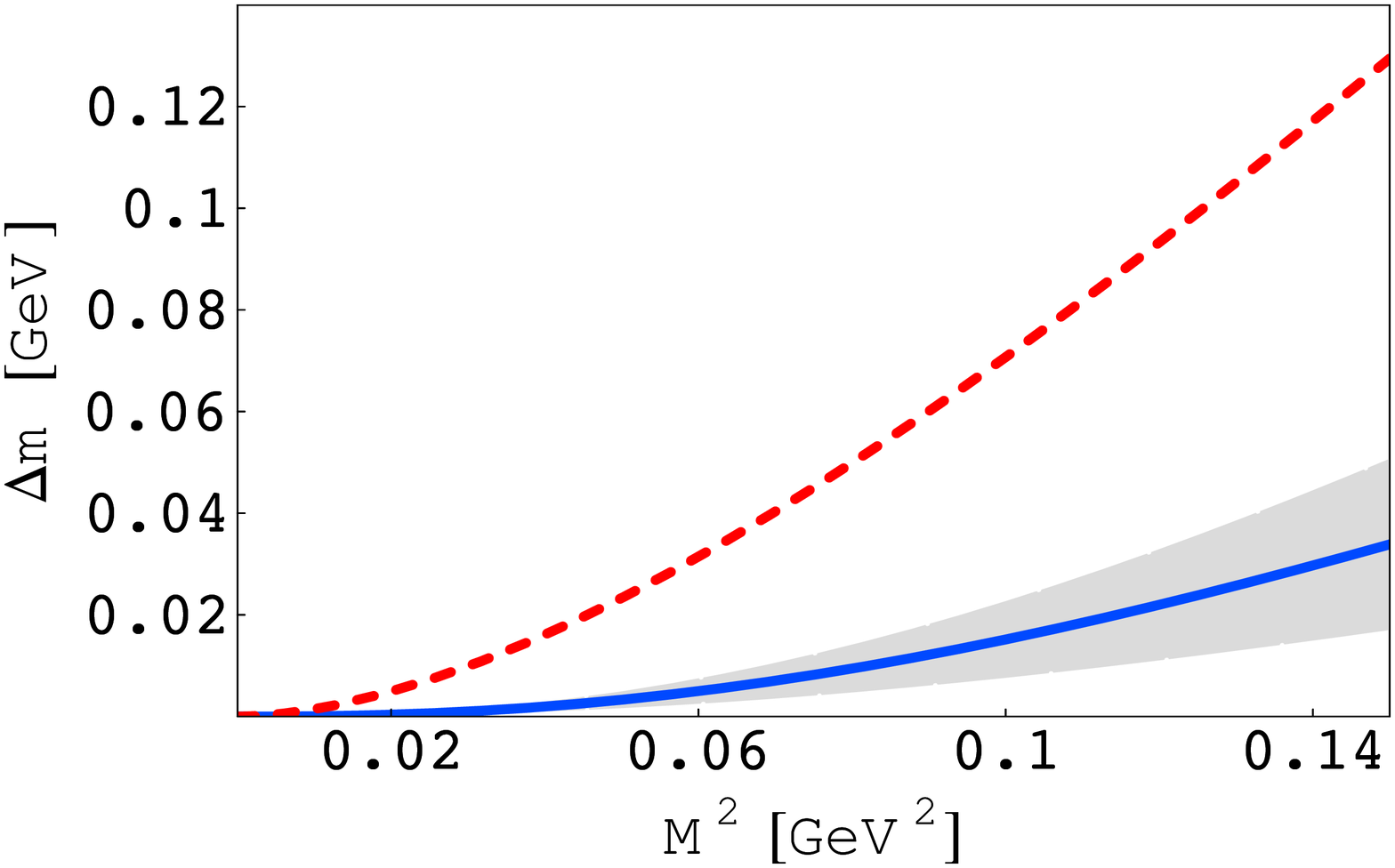}
}
\end{center}
\end{minipage}
\caption{Pion-mass dependence of terms contributing to the chiral expansion of the nucleon mass. Left panel: The solid line corresponds to $ k_5 M^5\logM$, the dashed line to $k_2M^3$. Right panel: The solid line corresponds to $ k_7 M^6\logMsq$, the dashed line to $k_3M^4\logM$. The grey band indicates an  error estimate (see Ref.~\protect\cite{Schindler:2007dr}).}\label{fig:NucMass}
\end{figure}
Contributions to the nucleon mass at $\calO (q^5)$, i.e., including two-loop diagrams, were first considered in Ref.~\cite{McGovern:1998tm}, and a complete calculation to  $\calO (q^6)$ was performed in Refs.~\cite{Schindler:2006ha,Schindler:2007dr}. The higher-order contributions take the form
\begin{equation}
\label{eq:NucMass6}
\begin{split}
m_N &= m + k_1 M^2 + k_2 M^3 + k_3 M^4 \logM +k_4 M^4 \\
&\quad + k_5 M^5\logM + k_6 M^5 + k_7 M^6 \logMsq + k_8 M^6 \logM + k_9 M^6.
\end{split}
\end{equation}
Since various so-far undetermined LECs enter the expressions for some of the higher-order $k_i$ it is not possible to give an accurate estimate of all terms in Eq.~(\ref{eq:NucMass6}). However, the fifth-order contribution $ k_5 M^5\logM$ is found to be $ k_5 M^5\ln\frac{M}{m_N} = -4.8\,\text{MeV}$ at the physical pion mass with $\mu=m_N$, while $ k_6 M^5 = 3.7 \,\text{MeV}$ or  $k_6 M^5 = -7.6\,\text{MeV}$ depending on the choice of the third-order LEC $d_{16}$ \cite{Schindler:2007dr}. Equation~(\ref{eq:NucMass6}) can also be used to examine the pion-mass dependence of the nucleon mass, which plays an important role in the extrapolation of lattice QCD to physical quark masses. Figure~\ref{fig:NucMass} shows the comparison of various terms in Eq.~(\ref{eq:NucMass6}) as a function of the squared pion mass. While the right panel shows the expected suppression of the higher-order term over the whole pion-mass range, the left panel indicates that the term $ k_5 M^5\logM$ becomes as large as $k_2M^3$ for a pion mass of roughly $M\sim 360\,\text{MeV}$. While this is not a reliable prediction of the behavior of higher-order contributions since only the leading nonanalytic parts are considered, the pion mass range at which the power counting is no longer applicable agrees with the estimates found in Refs.~\cite{Meissner:2005ba,Djukanovic:2006xc}.

\subsection{Virtual Compton scattering}

Real Compton scattering (RCS), $\gamma(q,\epsilon(\lambda))+N(p,s) \to \gamma(q',\epsilon'(\lambda'))+N(p',s')$, has a long history of providing important theoretical and experimental tests for models of nucleon structure (see, e.g., Refs. \cite{Lvov:1993fp,Scherer:1999yw,Drechsel:2002ar,Schumacher:2005an} for an introduction). Based on the requirement of gauge invariance, Lorentz invariance, crossing symmetry, and the discrete symmetries, the famous low-energy theorem of Low \cite{Low:1954kd} and Gell-Mann and Goldberger \cite{GellMann:1954kc} uniquely specifies the low-energy scattering amplitude up to and including terms linear in the photon momentum. The coefficients of this expansion are expressed in terms of global properties of the nucleon: its mass, charge, and anomalous magnetic moment $\kappa$. It is only terms of second order which contain new information on the structure of the nucleon specific to Compton scattering. For a general target, these effects can be parameterized in terms of two constants, the electric and magnetic polarizabilities $\alpha$ and $\beta$, respectively \cite{Klein:1955zz}.

As in all studies with electromagnetic probes, the possibilities to investigate the structure of the target are much greater if virtual photons are used, since energy and three-momentum of the virtual photon can be varied independently. Moreover, the longitudinal component of current operators entering the amplitude can be studied. The amplitude for virtual Compton scattering (VCS) off the proton, $T_{\rm VCS}$, is accessible in the reaction $e^-p\to e^-p\gamma$. Model-independent predictions, based on Lorentz invariance, gauge invariance, crossing symmetry, and the discrete symmetries, have been derived in Ref.\ \cite{Scherer:1996ux}. Up to and including terms of second order in the momenta $|\vec{q}\,|$ and $|\vec{q}\,'|$, the amplitude is completely specified in terms of quantities which can be obtained from elastic electron-proton scattering and real Compton scattering, namely $m_N$, $\kappa$, the electric and magnetic Sachs form factors $G_E$ and $G_M$, the electric mean square radius $r^2_E$, $\alpha_p$, and $\beta_p$. After dividing the amplitude $T_{\rm VCS}$ into a gauge-invariant generalized pole piece $T_{\rm pole}$ and a residual piece $T_{\rm R}$, the so-called generalized polarizabilities (GPs) of Ref.\ \cite{Guichon:1995pu} result from an analysis of the residual piece in terms of electromagnetic multipoles. A restriction to the lowest-order, i.e.\ terms linear in $\omega'$, leads to only electric and magnetic dipole radiation in the final state. Parity and angular-momentum selection rules, charge-conjugation symmetry, and particle crossing generate six independent GPs \cite{Guichon:1995pu,Drechsel:1996ag,Drechsel:1997xv}.

Predictions for the GPs of the nucleon have been obtained in HBChPT at ${\cal O}(q^3)$ \cite{Hemmert:1996gr,Hemmert:1997at} and ${\cal O}(q^4)$ \cite{Kao:2002cn,Kao:2004us}, as well as in the small-scale expansion\index{small-scale expansion} at ${\cal O}(q^3)$  \cite{Hemmert:1999pz}.  The predictions of HBChPT at ${\cal O}(q^3)$ contain no unknown LECs, i.e., they are given in terms of the pion mass, the axial-vector coupling constant, and the pion-decay constant. Table \ref{H1:b2:tableresults} shows a comparison between experimental results for the two structure functions $P_{LL}-P_{TT}/\epsilon$ and $P_{LT}$ at $Q^2=0.33$ GeV$^2$ obtained from a dedicated VCS experiment at MAMI \cite{Roche:2000ng} (see Ref.\ \cite{Janssens:2008qe} for an update) and the ${\cal O}(q^3)$ prediction \cite{Hemmert:1997at}. In view of the rather large value of $Q^2$ the agreement between experiment and HBChPT at ${\cal O}(q^3)$ is surprising and should be treated with care.

\begin{table*}
\begin{center}
\caption{Experimental results and theoretical predictions for the structure functions $P_{LL}-P_{TT}/\epsilon$ and $P_{LT}$ at $Q^2=0.33$ GeV$^2$ and $\epsilon=0.62$.} \label{H1:b2:tableresults}
\begin{tabular}{lcr}
\hline\noalign{\smallskip}&Experiment \cite{Roche:2000ng} &HBChPT \cite{Hemmert:1997at}\\
\hline\noalign{\smallskip}
$P_{LL}-P_{TT}/\epsilon$ $[\mbox{GeV}^{-2}]$ &
$23.7\pm 2.2_{\rm stat.}\pm 4.3_{\rm syst.} \pm 0.6_{\rm syst.norm.}$ &
26.0\\
$P_{LT}$ $[\mbox{GeV}^{-2}]$&
$-5.0\pm 0.8_{\rm stat.} \pm 1.4_{\rm syst.} \pm 1.1_{\rm syst. norm.}$ &
$-5.3$\\
\noalign{\smallskip}\hline
\end{tabular}
\end{center}
\end{table*}

\begin{figure}
\begin{center}
\resizebox{0.6\textwidth}{!}{%
\includegraphics{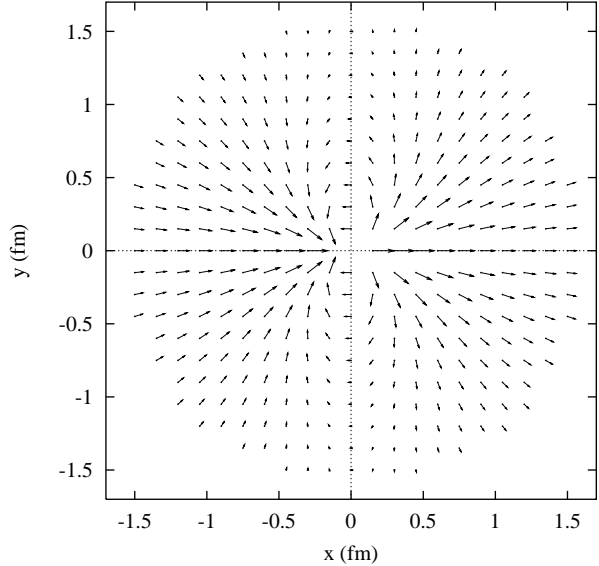}
}
\end{center}
\caption{Scaled electric polarization $r^3 \alpha_{i1}$ [10$^{-3}$ fm$^3$] \protect\cite{Lvov:2004}. The applied electric field points in the $x$ direction.} \label{fig:polarization}
\end{figure}

\begin{figure}
\begin{center}
\resizebox{0.45\textwidth}{!}{%
\includegraphics{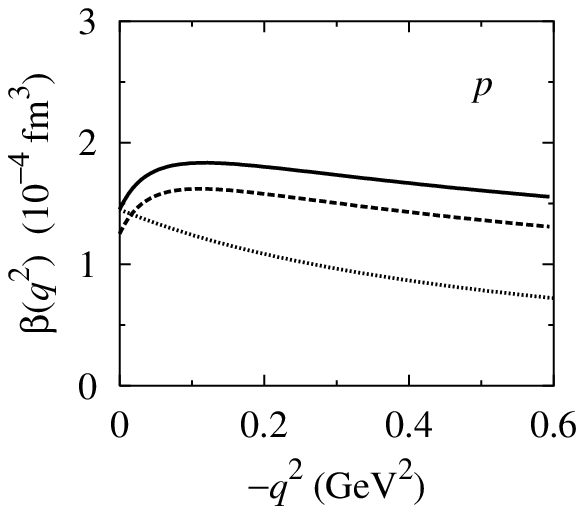}
}
\resizebox{0.45\textwidth}{!}{%
\includegraphics{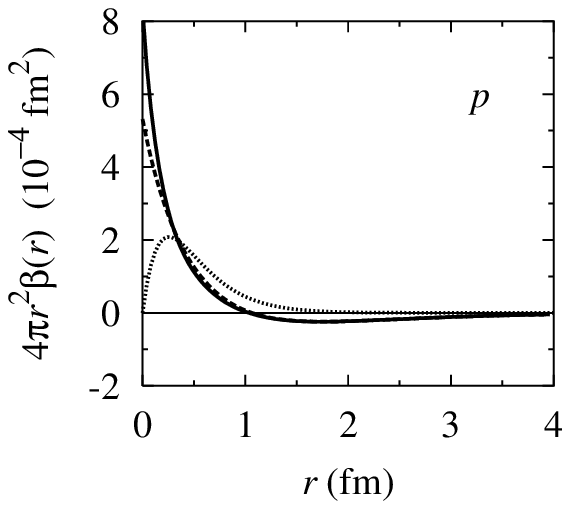}
}
\end{center}
\caption{Generalized magnetic polarizability $\beta(q^2)$ and density of magnetic polarizability $\beta(r)$ for the proton. Dashed lines: contribution of pion loops; solid lines: contribution of pion and kaon loops; dotted lines: vector meson dominance predictions normalized to ${\beta}(0)$ \protect\cite{L'vov:2001fz}.} \label{fig:beta}
\end{figure}

A covariant definition of the spin-averaged dipole polarizabilities was proposed in Ref.~\cite{L'vov:2001fz}. It was shown that {\it three} generalized dipole polarizabilities are needed to reconstruct spatial distributions. For example, if the nucleon is exposed to a static and uniform external electric field  $\vec{E}$, an electric polarization $\vec{\cal P}$ is generated which is related to the {\it density} of the induced electric dipole moments, 
\begin{equation}
\label{H1:b:d-induced} 
{\cal P}_i(\vec r) = 4\pi\alpha_{ij}(\vec r)\,E_j.
\end{equation}
The tensor $\alpha_{ij}(\vec r)$,  i.e.~the density of the full electric polarizability of the system, can be expressed as \cite{L'vov:2001fz} 
\begin{displaymath}
\alpha_{ij}(\vec r) =
\alpha_L(r) \hat r_i \hat r_j
+ \alpha_T(r) (\delta_{ij} - \hat r_i \hat r_j)
+ \frac{3\hat r_i \hat r_j - \delta_{ij}}{r^3}
 \int_r^\infty [\alpha_L(r')-\alpha_T(r')]\,r'^2\,dr',
\end{displaymath}
where $\alpha_L(r)$ and $\alpha_T(r)$ are Fourier transforms of the generalized longitudinal and transverse electric polarizabilities $\alpha_L(q^2)$ and $\alpha_T(q^2)$, respectively. In particular, it is important to realize that both longitudinal and transverse polarizabilities are needed to fully recover the electric polarization $\vec{\cal P}$. Figure \ref{fig:polarization} shows the induced polarization inside a proton as calculated in the framework of HBChPT at ${\cal O}(q^3)$ \cite{Lvov:2004}; the polarization, in general, does not point into the direction of the applied electric field. Similar considerations apply to an external magnetic field.
   Since the magnetic induction is always transverse (i.e., $\vec\nabla\cdot\vec B=0$), it is sufficient to consider $\beta_{ij}(\vec r)=\beta(r)\delta_{ij}$ \cite{L'vov:2001fz}.
  The induced magnetization $\vec{\cal M}$ is given in terms of the density of the magnetic polarizability as $\vec{\cal M}(\vec r) = 4\pi\beta(r)\vec B$ (see Fig.\ \ref{fig:beta}).

\subsection{Nucleon electromagnetic form factors}

\begin{figure}
\includegraphics[width=\textwidth]{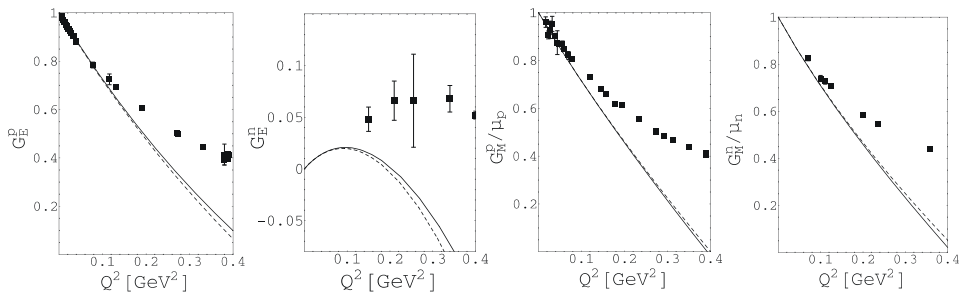}
\caption{Nucleon electromagnetic form factors in manifestly Lorentz-invariant BChPT up to and including $\calO (q^4)$. Solid line: EOMS. Dotted line: Infrared regularization (see Ref.~\protect\cite{Schindler:2005ke}). The experimental data are taken from Ref.~\protect\cite{Friedrich:2003iz}.}\label{fig:NucFF}
\end{figure}

The calculation of the nucleon electromagnetic form factors, which parameterize the matrix element of the electromagnetic current between one-nucleon states, presents a stringent test of BChPT due to the fact that the proton electromagnetic form factors are well-determined experimentally (see, e.g., Refs.~\cite{Perdrisat:2006hj,Drechsel:2007sq}).  They were calculated in HBChPT \cite{Bernard:1992qa}  and to $\calO (q^4)$ in infrared regularization \cite{Kubis:2000zd} and the EOMS scheme \cite{Fuchs:2003ir}. Baryon ChPT does not predict the nucleon radii, but uses them as input to determine a number of LECs. The results of  form factor calculations in manifestly Lorentz-invariant BChPT are shown in Fig.~\ref{fig:NucFF}. Agreement with the data breaks down around a momentum transfer of roughly $Q^2\approx 0.1\,\text{GeV}^2$, which is in agreement with the ``breakdown'' of the chiral expansion in Sec.~\ref{subsec:NucMass}. Clearly, a calculation to $\calO (q^4)$ does not produce sufficient curvature to also describe the data at higher $Q^2$. Additional, higher-order contributions therefore have to be taken into account to improve the agreement with experimental results. As explained above, a calculation beyond $\calO (q^4)$ in general also requires the inclusion of two-loop diagrams. Since the form factors in BChPT depend on two small scales, the pion mass and the momentum transfer, such calculations are highly complex. In addition, further unknown LECs appear at higher orders that have to be fixed by comparison with data. We therefore consider a different approach.

\section{Vector and axial-vector mesons in effective field theory}\label{sec:VM}

It has been well-known that vector mesons play an important role when considering the interaction of hadrons with electromagnetic fields.  For example, in the vector meson dominance model the hadrons couple to photons exclusively through intermediate vector mesons (see, e.g., Ref.~\cite{Sakurai}). In BChPT, vector mesons do not appear as explicit degrees of freedom. Instead their contributions are encoded in the values of the LECs. This can be seen by considering the expansion of a symbolical vector meson propagator \cite{Kubis:2000zd},
\begin{equation}
\label{eq:VMProp}
\frac{1}{q^2-M_V^2} = -\frac{1}{M_V^2} \left[ 1 + \frac{q^2}{M_V^2} + \left(\frac{q^2}{M_{V}^2}\right)^2+\ldots\right].
\end{equation}
Combined with the relevant coupling constants and numerical factors, each term on the right-hand side of Eq.~(\ref{eq:VMProp}) contributes to a LEC at a specific order. For a discussion of meson resonance exchange as a tool for estimating LECs see, e.g., Refs.~\cite{Ecker:1988te,Cirigliano:2006hb}. In order to account for all terms in the expansion it was suggested in Ref.~\cite{Kubis:2000zd} to keep vector mesons as explicit degrees of freedom. Since it was not known how to renormalize loop diagrams containing vector meson propagators, the authors of Ref.~\cite{Kubis:2000zd} set up a power counting that allowed them to only consider tree-level contributions.

\subsection{Nucleon electromagnetic form factors including vector mesons}\label{subsec:FFVM}

\begin{figure}
\includegraphics[width=\textwidth]{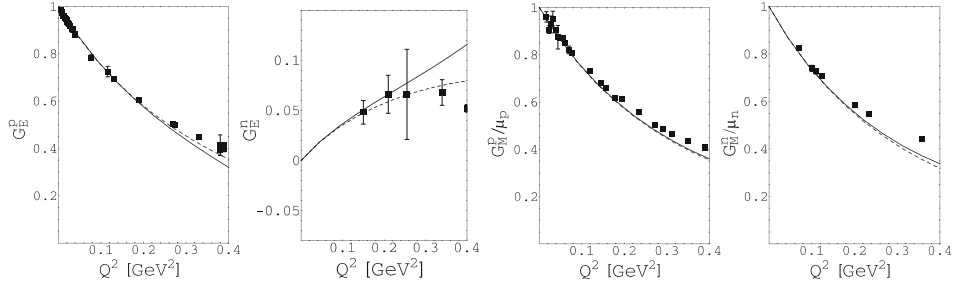}
\caption{Nucleon electromagnetic form factors in manifestly Lorentz-invariant BChPT with explicit $\rho$, $\omega$, and $\phi$ mesons up to and including $\calO (q^4)$. Solid line: EOMS. Dotted line: Infrared regularization (see Ref.~\protect\cite{Schindler:2005ke}). The experimental data are taken from Ref.~\protect\cite{Friedrich:2003iz}.}\label{fig:NucFFVM}
\end{figure}

With the EOMS scheme and the reformulation of infrared regularization two renormalization schemes were developed that are also applicable to loop diagrams containing vector mesons \cite{Fuchs:2003sh,Schindler:2003xv}. Both schemes were applied to the calculation of the electromagnetic nucleon form factors with explicit $\rho$, $\omega$, and $\phi$ mesons in Ref.~\cite{Schindler:2005ke}. It was shown that in infrared regularization to $\calO(q^4)$ all loop diagrams containing vector mesons vanish, which justifies the approach of Ref.~\cite{Kubis:2000zd}, while vector meson loop contributions are highly suppressed in the EOMS scheme. In both calculations the vector meson coupling were taken from dispersion relation analyses. The results for the form factors are shown in Fig.~\ref{fig:NucFFVM}.
They show an improved description of the data for larger momentum transfers up to $0.4\,\text{GeV}^2$ (and are in agreement with the improvements found in Ref.~\cite{Kubis:2000zd}). While this was expected on phenomenological grounds, it should be noted that the inclusion of vector mesons proceeds according to well-defined rules in an EFT formalism. Along similar lines, the impact of the axial-vector meson $a_1(1260)$ on the isovector axial form factor $G_A$ was discussed in Ref.~\cite{Schindler:2006it}.

\subsection{Universality of the $\rho$-meson coupling}

The universality of the $\rho$ meson coupling, i.e., the equality of the $\rho$ meson self-coupling and the coupling to nucleons and pions, plays an important role in vector meson dominance \cite{Sakurai}. Considering an effective Lagrangian describing the coupling of $\rho$ mesons to pions and nucleons, the relation $g_{\rho NN} = g$, with $g$ the $\rho$ self coupling, follows from chiral symmetry \cite{Weinberg:1968de}. On the other hand, the symmetries of the Lagrangian do not require the equality of the $\rho \pi\pi$ and $\rho NN$ couplings. However, it was shown in Ref.~\cite{Djukanovic:2004mm} that the universality of the $\rho$-meson coupling is a consequence of the requirement that the effective field theory describing $\rho$ mesons, pions, and nucleons is a consistent theory which can be renormalized. 

Starting from the Lagrangian $\calL$ of Ref.~\cite{Weinberg:1968de}, one can rewrite the Lagrangian in terms of renormalized fields and couplings, thereby introducing the basic Lagrangian
\begin{equation}
\label{eq:LagBasic}
\begin{split}
\calL_\text{basic} & = \frac{1}{2}\partial_\mu \pi^a\partial^\mu \pi^a
-\frac{M^{2}}{2}\pi^a \pi^a +\bar\Psi \left( i\gamma^\mu\partial_\mu - m \right)\Psi 
-\frac{1}{4}A^a_{\mu\nu} A^{a \mu\nu} +\frac {1}{2} M_{\rho}^2
\rho^a_{\mu} \rho^{a \mu}
\\
&+ g_{\rho\pi\pi}\epsilon^{abc}\pi^a\partial_\mu\pi^b \rho^{c\mu} -g\epsilon^{abc}\partial_\mu\rho_\nu^a \rho^{b\mu}\rho^{c\nu}
-\frac{1}{4} g^2 \epsilon^{abc}\epsilon^{ade} \rho^b_\mu
\rho^c_\nu \rho^{d \mu}\rho^{e \nu}  \\
&+g \bar\Psi\gamma^\mu \frac{\tau^a}{2}\Psi \rho^a_{\mu},
\end{split}
\end{equation}
with $A^a_{\mu\nu}\equiv\partial_\mu \rho^a_{\nu} -\partial_\nu \rho^a_{\mu}$, and the counter-term Lagrangian
\begin{equation}
\label{eq:LagCT}
\begin{split}
\calL_\text{ct}& =
-\frac{\delta Z_\rho}{4}\ A^a_{\mu\nu}A^{a \mu\nu}
+\left[\delta g_{\rho\pi\pi}
+g_{\rho\pi\pi}\left(\frac{\delta Z_\rho}{2}+\delta Z_\pi\right)\right]
\epsilon^{abc}\pi^a\partial_\mu\pi^b \rho^{c\mu}  \\
&-\left(\delta g+\frac{3}{2} g \delta Z_\rho\right)
\epsilon^{abc}\partial_\mu\rho_\nu^a \rho^{b\mu}\rho^{c\nu}  +\left[\delta g+g\left(\frac{\delta Z_\rho}{2}+ \delta Z_\Psi\right)\right]
\bar\Psi\gamma^\mu \frac{\tau^a}{2}\Psi \rho^a_{\mu},
\end{split}
\end{equation}
with 
\begin{equation}
\calL  = \calL_\text{basic}+\calL_\text{ct} + \tilde{\calL}_1,
\end{equation}
where $\tilde{\calL}_1$ is a residual Lagrangian containing all higher-dimensional interactions and all remaining counter terms. The authors of Ref.~\cite{Djukanovic:2004mm} considered the $\rho$-meson self energy as well as the $\rho\rho\rho$ and $\rho\bar\Psi\Psi$ vertex functions to one-loop order. These receive contributions from one-loop diagrams containing the renormalized couplings of Eq.~(\ref{eq:LagBasic}) and from tree-level diagrams with one-loop order counter terms of Eq.~(\ref{eq:LagCT}). Requiring that the results are UV finite introduces relations between the couplings of the theory, resulting in
\begin{equation}
\label{eq:CoupRel}
g_{\rho\pi\pi}^3=g g_{\rho\pi\pi}^2.
\end{equation}
Equation (\ref{eq:CoupRel}) has two solutions: the trivial one $g_{\rho\pi\pi} = 0$, corresponding to a theory in which the $\rho$ meson does not couple to pions, and the non-trivial one
\begin{equation}
\label{eq:Univ}
g_{\rho\pi\pi}=g.
\end{equation}
Universality is therefore the {\it consequence}  of the existence of a consistent EFT with $\rho$ mesons, pions, and nucleons. From Eq.~(\ref{eq:Univ}) one can also derive the well-known Kawarabayashi-Suzuki-Riazuddin-Fayyazuddin relation \cite{Kawarabayashi:1966kd,Riazuddin:sw},
\begin{equation}
g^2=\frac{M_{\rho }^2}  {2 F^2}. \label{ksrf}
\end{equation}

In Ref.~\cite{Djukanovic:2005ag}, the same authors considered the extension of this approach to also include the interactions with photons. Again requiring UV-finite expressions at one-loop order results in relations between various couplings, which in turn determine the $\rho^+$ gyromagnetic ratio to have a value of $2$ and the $\rho$ meson mass splitting to be
\begin{equation}
M_{\rho^0}-M_{\rho^+} \sim 1\,\text{MeV},
\end{equation}
which has to be compared with the PDG value $(-0.7\pm 0.8)$ MeV \cite{Nakamura:2010zzi}.
Finally, it was shown in Ref.~\cite{Djukanovic:2010tb} that the massive Yang-Mills part of Eq.~(\ref{eq:LagBasic}) for the $\rho$ mesons is indeed the most general leading-order Lagrangian once self consistency in the sense of constraints and perturbative renormalizability is imposed.

\subsection{Complex-mass scheme and effective field theory}

   In Sec.\ \ref{subsec:FFVM} we saw how the inclusion of virtual vector mesons generates an improved description of the electromagnetic form factors, for which ordinary chiral perturbation theory does not produce sufficient curvature. 
   So far the inclusion of virtual vector mesons has been restricted to low-energy processes in which the vector mesons cannot be generated explicitly.
   However, one would also like to investigate the properties of hadronic resonances such as their masses and widths as well as their electromagnetic properties. 
   An extension of chiral effective field theory to the momentum region near the complex pole corresponding to the vector mesons was proposed in Ref.~\cite{Djukanovic:2009zn}, in which the power-counting problem was addressed by applying the complex-mass scheme (CMS) \cite{Stuart:1990,Denner:1999gp,Denner:2005fg,Denner:2006ic,Actis:2006rc,Actis:2008uh} to the effective field theory.
   Since the $\rho$ mass is not treated as a small quantity, the presence of large external four-momenta, e.g., in terms of the zeroth component, leads to a considerable complication regarding the power counting of loop diagrams.
   To assign a chiral order to a given diagram it is first necessary to investigate all possibilities how the external momenta could flow through the internal lines of that diagram.
   Next, when assigning powers to propagators and vertices, one needs to determine the chiral order for a given flow of external momenta.
   Finally, the smallest order resulting from the various assignments is defined as the chiral order of the given diagram.
   The application of the CMS to the renormalization of loop diagrams amounts to splitting the bare parameters of the Lagrangian into renormalized parameters and counter terms and choosing the renormalized masses as the {\it complex} poles of the dressed propagators in the chiral limit, $M_R^2=(M_\chi-i\Gamma_\chi/2)^2$.
   The result for the chiral expansion of the pole mass and the width of
the $\rho$ meson to ${\cal O}(q^4)$ reads \cite{Djukanovic:2009zn}
\begin{eqnarray}
M_\rho^2  & = & M_\chi^2 +c_x M^2
-\frac{g_{\omega\rho\pi}^2 M^3 M_\chi}{24 \pi }\nonumber\\
&&+\frac{M^4}{32\pi^2 F^2} \left(3 -2\, \ln \frac{M^2}{M_\chi^2}\right) +\frac{g_{\omega\rho\pi}^2 M^4}{32 \pi
^2} \left(1-\ln \frac{M^2}{M_\chi^2}\right)\,,
\label{phmass}\\
\Gamma & = & \Gamma_\chi +\frac{\Gamma_\chi ^3}{8 M_\chi^2}-\frac{c_x \Gamma_\chi  M^2}{2 M_\chi^2}
-\frac{g_{\omega\rho\pi}^2 M^3 \Gamma_\chi}{48 \pi \,M_\chi} +\frac{ M^4}{16\,\pi \,F^2 M_\chi}\,.
\label{phwidth}
\end{eqnarray}
  Here, $M^2$ is the lowest-order expression for the squared pion mass, $F$ the pion-decay constant in the chiral limit, $c_x$ a low-energy coupling constant of the $\pi\rho$ Lagrangian, and $g_{\omega\rho\pi}$ a coupling constant.
   The nonanalytic terms of Eq.~(\ref{phmass}) agree with the results of Ref.~\cite{Leinweber:2001ac}.
   Both mass $M_\chi$ and width $\Gamma_\chi$ in the chiral limit are input parameters in this approach. The numerical importance of the different contributions has been estimated using
$$
F=0.092 \,{\rm GeV},\quad M=0.139 \,{\rm GeV}\,,\quad g_{\omega\rho\pi} = 16 \,{\rm GeV^{-1}},\quad M_\chi\approx M_\rho=0.78\,{\rm GeV},
$$
resulting in the expansion (units of GeV$^2$ and GeV, respectively)
\begin{eqnarray}
M_\rho^2 & = & M_\chi^2+0.019 \,c_x - 0.0071 + 0.0014 +0.0013\,,\nonumber\\
\Gamma & \approx & \Gamma_\chi + 0.21\,\Gamma_\chi^3-0.016\,c_x \Gamma_\chi -0.0058\,\Gamma_\chi + 0.0011\,.
\label{numestimate}
\end{eqnarray}
For pion masses larger than $M_\rho/2$ the $\rho$ meson becomes a stable particle. For such values of the pion mass the series of Eq.~(\ref{phwidth}) should diverge. Along similar lines, Ref.~\cite{Djukanovic:2009gt} contains a calculation of the mass and the width of the Roper resonance using the CMS.

\section{Conclusions}

The (approximate) chiral symmetry and its spontaneous breaking are two important properties of QCD for low-energy processes. The effective field theories presented here allow a systematic study of the consequences of these QCD features. We have only presented a small snapshot of the vast number and successes of the application of chiral effective field theories. In addition to the meson and one-nucleon sectors discussed here, an extensive research effort is focused on the inclusion of the $\Delta$ resonance (see, e.g., Refs.~\cite{Hemmert:1997ye,Hacker:2005fh,Pascalutsa:2006up}) as well as applying these techniques to two- and few-nucleon systems (see, e.g., Ref.~\cite{Epelbaum:2008ga} and references therein). All these EFT calculations have in common that they are model-independent and systematically improvable approximations to QCD in the low-energy domain. As briefly discussed in Sec.~\ref{subsec:NucMass}, chiral perturbation theory also presents an important tool in the extrapolation of lattice QCD results. In addition to the quark mass expansion, effective field theory techniques have been developed to treat finite-volume effects, as well as different formulations of the lattice action (see, e.g., Ref.~\cite{Golterman:2009kw} and references therein).

\begin{acknowledgement}
We are grateful to the organizers of the conclusive symposium ``Many-body structure of strongly interacting systems'' for a very interesting and enjoyable meeting. We would like to thank our collaborators for productive joint efforts and interesting discussions. Special thanks go to J.~Gegelia for a long-standing and fruitful collaboration. Much of this work would not have been possible without the support of the Collaborative Research Center 443 (SFB 443) of the Deutsche Forschungsgemeinschaft.
\end{acknowledgement}


\begin{thebibliography}{}

\bibitem{Polchinski:1992ed}
  J.~Polchinski,
  arXiv:hep-th/9210046

\bibitem{Georgi:1994qn}
  H.~Georgi,
  Ann.\ Rev.\ Nucl.\ Part.\ Sci.\  {\bf 43} (1993) 209

\bibitem{Kaplan:1995uv}
D.~B.~Kaplan,
arXiv:nucl-th/9506035

\bibitem{Manohar:1996cq}
A.~V.~Manohar,
arXiv:hep-ph/9606222

\bibitem{Pich:1998xt}
A.~Pich,
arXiv:hep-ph/9806303

\bibitem{Burgess:1998ku}
  C.~P.~Burgess,
  Phys.\ Rept.\  {\bf 330} (2000) 193

\bibitem{Ecker:2005ny}
  G.~Ecker,
  arXiv:hep-ph/0507056

\bibitem{Epelbaum:2005pn}
  E.~Epelbaum,
  Prog.\ Part.\ Nucl.\ Phys.\  {\bf 57} (2006) 654

\bibitem{Kaplan:2005es}
  D.~B.~Kaplan,
  arXiv:nucl-th/0510023

\bibitem{Donoghue:2009mn}
  J.~F.~Donoghue,
  PoS {\bf EFT09} (2009) 001
  
\bibitem{Schindler:2011LNP}
M.~R.~Schindler, S.~Scherer,
to appear in Lect.\ Notes Phys.\  {\bf 830} (2011) 1

\bibitem{Weinberg:1978kz}
  S.~Weinberg,
  Physica A {\bf 96} (1979) 327

\bibitem{Gasser:1983yg}
  J.~Gasser, H.~Leutwyler,
  Annals Phys.\  {\bf 158} (1984) 142

\bibitem{Gasser:1984gg}
  J.~Gasser, H.~Leutwyler,
  Nucl.\ Phys.\  B {\bf 250} (1985) 465

\bibitem{Fearing:1994ga}
  H.~W.~Fearing, S.~Scherer,
  Phys.\ Rev.\  D {\bf 53} (1996) 315

\bibitem{Bijnens:1999sh}
  J.~Bijnens, G.~Colangelo, G.~Ecker,
  JHEP {\bf 9902} (1999) 020

\bibitem{Ebertshauser:2001nj}
T.~Ebertsh\"auser, H.~W.~Fearing, S.~Scherer,
  Phys.\ Rev.\  D {\bf 65} (2002) 054033

\bibitem{Bijnens:2001bb}
  J.~Bijnens, L.~Girlanda, P.~Talavera,
  Eur.\ Phys.\ J.\  C {\bf 23} (2002) 539

\bibitem{Meissner:1993ah}
  U.-G.~Mei{\ss}ner,
  Rept.\ Prog.\ Phys.\  {\bf 56} (1993) 903

\bibitem{Leutwyler:1994fi}
  H.~Leutwyler,
  arXiv:hep-ph/9406283

\bibitem{Ecker:1994gg}
  G.~Ecker,
  Prog.\ Part.\ Nucl.\ Phys.\  {\bf 35} (1995) 1

\bibitem{Pich:1995bw}
  A.~Pich,
  Rept.\ Prog.\ Phys.\  {\bf 58} (1995) 563

\bibitem{Scherer:2002tk}
S.~Scherer,
Adv.\ Nucl.\ Phys.\  {\bf 27} (2003) 277

\bibitem{Bijnens:2006zp}
J.~Bijnens,
Prog.\ Part.\ Nucl.\ Phys.\  {\bf 58} (2007) 521

\bibitem{Li:1971vr}
L.~F.~Li, H.~Pagels,
Phys.\ Rev.\ Lett.\  {\bf 26} (1971) 1204

\bibitem{Bijnens:1987dc}
J.~Bijnens, F.~Cornet,
Nucl.\ Phys.\ B {\bf 296} (1988) 557

\bibitem{Terentev:1972ix}
M.~V.~Terent'ev,
  Sov.\ J.\ Nucl.\ Phys.\  {\bf 16} (1973) 87
  [Yad.\ Fiz.\  {\bf 16} (1972) 162]

\bibitem{Frlez:2003pe}
E.~Frle{\v z} {\it et al.},
Phys.\ Rev.\ Lett.\  {\bf 93} (2004) 181804

\bibitem{Burgi:1996qi}
U.~B\"urgi,
Nucl.\ Phys.\ B {\bf 479} (1996) 392

\bibitem{Gasser:2006qa}
J.~Gasser, M.~A.~Ivanov, M.~E.~Sainio,
Nucl.\ Phys.\ B {\bf 745}  (2006) 84

\bibitem{Bijnens:1995yn}
J.~Bijnens, G.~Colangelo, G.~Ecker, J.~Gasser, M.~E.~Sainio,
Phys.\ Lett.\ B {\bf 374} (1996) 210

\bibitem{Antipov:1982kz}
Y.~M.~Antipov {\it et al.},
Phys.\ Lett.\ B {\bf 121} (1983) 445

\bibitem{Aibergenov:1986gi}
T.~A.~Aibergenov {\it et al.},
Czech.\ J.\ Phys.\ B {\bf 36} (1986) 948

\bibitem{Ahrens:2004mg}
J.~Ahrens {\it et al.},
Eur.\ Phys.\ J.\ A {\bf 23} (2005) 113

\bibitem{Berger:1984xb}
PLUTO Collaboration (C.~Berger {\it et al.}),
Z.\ Phys.\ C {\bf 26} (1984) 199

\bibitem{Courau:1986gn}
DM1 Collaboration (A.~Courau {\it et al.}),
Nucl.\ Phys.\ B {\bf 271} (1986) 1

\bibitem{Ajaltoni}
DM2 Collaboration (Z.~Ajaltoni {\it et al.}), in {\it Proceedings of the VII
International Workshop on Photon-Photon Collisions, Paris, 1-5 April 1986},
edited by A.~Courau, P.~Kessler (World Scientific, Singapore, 1986)

\bibitem{Boyer:1990vu}
MARK II Collaboration (J.~Boyer {\it et al.}),
Phys.\ Rev.\ D {\bf 42} (1990) 1350

\bibitem{Guskov:2008zz}
A.~Guskov on behalf of the COMPASS collaboration,
J.\ Phys.\ Conf.\ Ser.\  {\bf 110} (2008) 022016

\bibitem{Drechsel:1994kh}
D.~Drechsel, L.~V.~Fil'kov,
Z.\ Phys.\ A {\bf 349} (1994) 177

\bibitem{Chew:1958wd}
G.~F.~Chew, F.~E.~Low,
Phys.\ Rev.\  {\bf 113} (1959) 1640

\bibitem{Unkmeir}
C.~Unkmeir, PhD thesis, Johannes Gutenberg-Universit\"at, Mainz (2000)

\bibitem{Kao:2008pf}
  C.~W.~Kao, B.~E.~Norum, K.~Wang,
  Phys.\ Rev.\  D {\bf 79} (2009) 054001

\bibitem{Fil'kov:1998np}
L.~V.~Fil'kov, V.~L.~Kashevarov,
Eur.\ Phys.\ J.\ A {\bf 5} (1999) 285

\bibitem{Fil'kov:2005ss}
L.~V.~Fil'kov, V.~L.~Kashevarov,
Phys.\ Rev.\  C {\bf 73} (2006) 035210

\bibitem{Pasquini:2008ep}
B.~Pasquini, D.~Drechsel, S.~Scherer,
Phys.\ Rev.\  C {\bf 77} (2008) 065211

\bibitem{Unkmeir:1999md}
C.~Unkmeir, S.~Scherer, A.~I.~L'vov, D.~Drechsel,
Phys.\ Rev.\ D {\bf 61} (2000) 034002

\bibitem{Fuchs:2000pn}
T.~Fuchs, B.~Pasquini, C.~Unkmeir S.~Scherer,
Czech.\ J.\ Phys.\  {\bf 52} (2002) B135

\bibitem{L'vov:2001fz}
A.~I.~L'vov, S.~Scherer, B.~Pasquini, C.~Unkmeir, D.~Drechsel,
Phys.\ Rev.\ C {\bf 64} (2001) 015203

\bibitem{Unkmeir:2001gw}
C.~Unkmeir, A.~Ocherashvili, T.~Fuchs, M.~A.~Moinester, S.~Scherer,
Phys.\ Rev.\ C {\bf 65} (2002) 015206

\bibitem{Gasser:1987rb}
  J.~Gasser, M.~E.~Sainio, A.~\v{S}varc,
  Nucl.\ Phys.\  B {\bf 307} (1988) 779
  
\bibitem{Krause:1990xc}
A.~Krause,
Helv.\ Phys.\ Acta {\bf 63} (1990) 3
  
\bibitem{Fettes:2000gb}
  N.~Fettes, U.-G.~Mei{\ss}ner, M.~Moj\v{z}i\v{s}, S.~Steininger,
  Annals Phys.\  {\bf 283} (2000) 273
  [Erratum-ibid.\  {\bf 288} (2001) 249]
  
\bibitem{Jenkins:1990jv}
  E.~E.~Jenkins, A.~V.~Manohar,
  Phys.\ Lett.\  B {\bf 255} (1991) 558
  
\bibitem{Ellis:1997kc}
  P.~J.~Ellis, H.~B.~Tang,
  Phys.\ Rev.\  C {\bf 57} (1998) 3356

\bibitem{Becher:1999he}
  T.~Becher, H.~Leutwyler,
  Eur.\ Phys.\ J.\  C {\bf 9} (1999) 643
  
\bibitem{Gegelia:1999gf}
  J.~Gegelia, G.~Japaridze,
  Phys.\ Rev.\  D {\bf 60} (1999) 114038

\bibitem{Gegelia:1999qt}
  J.~Gegelia, G.~Japaridze, X.~Q.~Wang,
  J.\ Phys.\ G {\bf 29} (2003) 2303

\bibitem{Fuchs:2003qc}
  T.~Fuchs, J.~Gegelia, G.~Japaridze, S.~Scherer,
  Phys.\ Rev.\  D {\bf 68} (2003) 056005
  
\bibitem{Bernard:1995dp}
  V.~Bernard, N.~Kaiser, U.-G.~Mei{\ss}ner,
  Int.\ J.\ Mod.\ Phys.\  E {\bf 4} (1995) 193

\bibitem{Bernard:2007zu}
  V.~Bernard,
  Prog.\ Part.\ Nucl.\ Phys.\  {\bf 60} (2008) 82

\bibitem{Schindler:2003je}
  M.~R.~Schindler, J.~Gegelia, S.~Scherer,
  Nucl.\ Phys.\  B {\bf 682} (2004) 367

\bibitem{Fuchs:2003sh}
  T.~Fuchs, M.~R.~Schindler, J.~Gegelia, S.~Scherer,
  Phys.\ Lett.\  B {\bf 575} (2003) 11

\bibitem{Schindler:2003xv}
  M.~R.~Schindler, J.~Gegelia, S.~Scherer,
  Phys.\ Lett.\  B {\bf 586} (2004) 258

\bibitem{Bruns:2004tj}
  P.~C.~Bruns, U.-G.~Mei{\ss}ner,
  Eur.\ Phys.\ J.\  C {\bf 40} (2005) 97
  
\bibitem{Scherer:2009bt}
  S.~Scherer,
  Prog.\ Part.\ Nucl.\ Phys.\  {\bf 64} (2010) 1
  
\bibitem{Bernard:1992qa}
  V.~Bernard, N.~Kaiser, J.~Kambor, U.-G.~Mei{\ss}ner,
  Nucl.\ Phys.\  B {\bf 388} (1992) 315
  
\bibitem{McGovern:1998tm}
  J.~A.~McGovern, M.~C.~Birse,
  Phys.\ Lett.\  B {\bf 446} (1999) 300
  
\bibitem{Fuchs:2003kq}
  T.~Fuchs, J.~Gegelia, S.~Scherer,
  Eur.\ Phys.\ J.\  A {\bf 19} (2004) 35

\bibitem{Schindler:2006ha}
  M.~R.~Schindler, D.~Djukanovic, J.~Gegelia, S.~Scherer,
  Phys.\ Lett.\  B {\bf 649} (2007) 390

\bibitem{Schindler:2007dr}
  M.~R.~Schindler, D.~Djukanovic, J.~Gegelia, S.~Scherer,
  Nucl.\ Phys.\  A {\bf 803} (2008) 68

\bibitem{Meissner:2005ba}
  U.-G.~Mei{\ss}ner,
  PoS {\bf LAT2005} (2006) 009

\bibitem{Djukanovic:2006xc}
  D.~Djukanovic, J.~Gegelia, S.~Scherer,
  Eur.\ Phys.\ J.\  A {\bf 29} (2006) 337

\bibitem{Lvov:1993fp}
A.~I.~L'vov,
Int.\ J.\ Mod.\ Phys.\  A {\bf 8} (1993) 5267

\bibitem{Scherer:1999yw}
S.~Scherer,
Czech.\ J.\ Phys.\  {\bf 49} (1999) 1307

\bibitem{Drechsel:2002ar}
D.~Drechsel, B.~Pasquini, M.~Vanderhaeghen,
Phys.\ Rept.\  {\bf 378} (2003) 99

\bibitem{Schumacher:2005an}
M.~Schumacher,
Prog.\ Part.\ Nucl.\ Phys.\  {\bf 55} (2005) 567

\bibitem{Low:1954kd}
F.~E.~Low,
Phys.\ Rev.\  {\bf 96} (1954) 1428

\bibitem{GellMann:1954kc}
M.~Gell-Mann, M.~L.~Goldberger,
Phys.\ Rev.\  {\bf 96} (1954) 1433

\bibitem{Klein:1955zz}
A.~Klein,
Phys.\ Rev.\  {\bf 99} (1955) 998

\bibitem{Scherer:1996ux}
S.~Scherer, A.~Y.~Korchin, J.~H.~Koch,
Phys.\ Rev.\  C {\bf 54} (1996) 904

\bibitem{Guichon:1995pu}
P.~A.~M.~Guichon, G.~Q.~Liu, A.~W.~Thomas,
Nucl.\ Phys.\  A {\bf 591} (1995) 606

\bibitem{Drechsel:1996ag}
D.~Drechsel, G.~Kn\"ochlein, A.~Metz, S.~Scherer,
Phys.\ Rev.\  C {\bf 55} (1997) 424

\bibitem{Drechsel:1997xv}
D.~Drechsel, G.~Kn\"ochlein, A.~Y.~Korchin, A.~Metz, S.~Scherer,
Phys.\ Rev.\  C {\bf 57} (1998) 941

\bibitem{Hemmert:1996gr}
T.~R.~Hemmert, B.~R.~Holstein, G.~Kn\"ochlein, S.~Scherer,
Phys.\ Rev.\  D {\bf 55} (1997) 2630

\bibitem{Hemmert:1997at}
T.~R.~Hemmert, B.~R.~Holstein, G.~Kn\"ochlein, S.~Scherer,
Phys.\ Rev.\ Lett.\  {\bf 79} (1997) 22

\bibitem{Kao:2002cn}
C.~W.~Kao, M.~Vanderhaeghen,
Phys.\ Rev.\ Lett.\  {\bf 89} (2002) 272002

\bibitem{Kao:2004us}
C.~W.~Kao, B.~Pasquini, M.~Vanderhaeghen,
Phys.\ Rev.\  D {\bf 70} (2004) 114004

\bibitem{Hemmert:1999pz}
T.~R.~Hemmert, B.~R.~Holstein, G.~Kn\"ochlein, D.~Drechsel,
Phys.\ Rev.\  D {\bf 62} (2000) 014013

\bibitem{Roche:2000ng}
J.~Roche {\it et al.}  [VCS Collaboration and A1 Collaboration],
Phys.\ Rev.\ Lett.\  {\bf 85} (2000) 708

\bibitem{Janssens:2008qe}
P.~Janssens {\it et al.}  [A1 Collaboration],
Eur.\ Phys.\ J.\  A {\bf 37} (2008) 1

\bibitem{Lvov:2004}
A.~I.~L'vov, S.~Scherer, unpublished.

\bibitem{Perdrisat:2006hj}
  C.~F.~Perdrisat, V.~Punjabi, M.~Vanderhaeghen,
  Prog.\ Part.\ Nucl.\ Phys.\  {\bf 59} (2007)  694
  
\bibitem{Drechsel:2007sq}
  D.~Drechsel, T.~Walcher,
  Rev.\ Mod.\ Phys.\  {\bf 80} (2008) 731
  
\bibitem{Kubis:2000zd}
  B.~Kubis, U.-G.~Mei{\ss}ner,
  Nucl.\ Phys.\  A {\bf 679} (2001) 698

\bibitem{Fuchs:2003ir}
  T.~Fuchs, J.~Gegelia, S.~Scherer,
  J.\ Phys.\ G {\bf 30} (2004) 1407

\bibitem{Friedrich:2003iz}
  J.~Friedrich, T.~Walcher,
  Eur.\ Phys.\ J.\  A {\bf 17} (2003) 607

\bibitem{Sakurai}
J.~J.~Sakurai, \textit{Currents and mesons} (The University of Chicago Press, Chicago 1969)

\bibitem{Ecker:1988te}
  G.~Ecker, J.~Gasser, A.~Pich, E.~de Rafael,
  Nucl.\ Phys.\  B {\bf 321} (1989) 311

\bibitem{Cirigliano:2006hb}
  V.~Cirigliano, G.~Ecker, M.~Eidem\"uller, R.~Kaiser, A.~Pich, J.~Portol\'es,
  Nucl.\ Phys.\  B {\bf 753} (2006) 139

\bibitem{Schindler:2005ke}
  M.~R.~Schindler, J.~Gegelia, S.~Scherer,
  Eur.\ Phys.\ J.\  A {\bf 26} (2005) 1

\bibitem{Schindler:2006it}
  M.~R.~Schindler, T.~Fuchs, J.~Gegelia, S.~Scherer,
  Phys.\ Rev.\  C {\bf 75} (2007) 025202

\bibitem{Weinberg:1968de}
  S.~Weinberg,
  Phys.\ Rev.\  {\bf 166} (1968) 1568

\bibitem{Djukanovic:2004mm}
  D.~Djukanovic, M.~R.~Schindler, J.~Gegelia, G.~Japaridze, S.~Scherer,
  Phys.\ Rev.\ Lett.\  {\bf 93} (2004) 122002

\bibitem{Kawarabayashi:1966kd}
K.~Kawarabayashi and M.~Suzuki,
  Phys.\ Rev.\ Lett.\  {\bf 16} (1966) 255

\bibitem{Riazuddin:sw}
Riazuddin and Fayyazuddin,
  Phys.\ Rev.\  {\bf 147} (1966) 1071

\bibitem{Djukanovic:2005ag}
  D.~Djukanovic, M.~R.~Schindler, J.~Gegelia, S.~Scherer,
  Phys.\ Rev.\ Lett.\  {\bf 95} (2005) 012001

\bibitem{Nakamura:2010zzi}
  K.~Nakamura {\it et al.} (Particle Data Group),
  J.\ Phys.\ G {\bf 37 } (2010)  075021

\bibitem{Djukanovic:2010tb}
  D.~Djukanovic, J.~Gegelia, S.~Scherer,
  Int.\ J.\ Mod.\ Phys.\  A {\bf 25} (2010) 3603

\bibitem{Djukanovic:2009zn}
  D.~Djukanovic, J.~Gegelia, A.~Keller, S.~Scherer,
  Phys.\ Lett.\  B {\bf 680} (2009) 235

\bibitem{Stuart:1990}
R.~G.~Stuart, in ${\rm Z}^0$ {\it Physics}, ed. J. Tran Thanh Van
(Editions Frontieres, Gif-sur-Yvette, 1990), p.~41

\bibitem{Denner:1999gp}
A.~Denner, S.~Dittmaier, M.~Roth, D.~Wackeroth,
Nucl.\ Phys.\  B {\bf 560} (1999) 33

\bibitem{Denner:2005fg}
  A.~Denner, S.~Dittmaier, M.~Roth, L.~H.~Wieders,
  Nucl.\ Phys.\  B {\bf 724} (2005) 247

\bibitem{Denner:2006ic}
  A.~Denner, S.~Dittmaier,
  Nucl.\ Phys.\ Proc.\ Suppl.\  {\bf 160} (2006) 22

\bibitem{Actis:2006rc}
  S.~Actis, G.~Passarino,
  Nucl.\ Phys.\  B {\bf 777} (2007) 100

\bibitem{Actis:2008uh}
  S.~Actis, G.~Passarino, C.~Sturm, S.~Uccirati,
  Phys.\ Lett.\  B {\bf 669} (2008) 62

\bibitem{Leinweber:2001ac}
  D.~B.~Leinweber, A.~W.~Thomas, K.~Tsushima, S.~V.~Wright,
  Phys.\ Rev.\  D {\bf 64} (2001) 094502

\bibitem{Djukanovic:2009gt}
D.~Djukanovic, J.~Gegelia, S.~Scherer,
Phys.\ Lett.\ B {\bf 690} (2010) 123

\bibitem{Hemmert:1997ye}
  T.~R.~Hemmert, B.~R.~Holstein, J.~Kambor,
  J.\ Phys.\ G {\bf 24} (1998) 1831

\bibitem{Hacker:2005fh}
  C.~Hacker, N.~Wies, J.~Gegelia, S.~Scherer,
  Phys.\ Rev.\  C {\bf 72} (2005) 055203

\bibitem{Pascalutsa:2006up}
  V.~Pascalutsa, M.~Vanderhaeghen, S.~N.~Yang,
  Phys.\ Rept.\  {\bf 437} (2007) 125


\bibitem{Epelbaum:2008ga}
  E.~Epelbaum, H.~W.~Hammer, U.-G.~Mei{\ss}ner,
  Rev.\ Mod.\ Phys.\  {\bf 81} (2009) 1773

\bibitem{Golterman:2009kw}
  M.~Golterman,
arXiv:0912.4042 [hep-lat]
  
\end{thebibliography}
\end{document}